\documentclass[final,5p,times,twocolumn,numeric]{elsarticle}

\usepackage{amssymb}
\usepackage[dvipsnames]{xcolor}
\usepackage{amsthm}
\usepackage{mathtools}
\usepackage{abraces}
\usepackage{moreverb,url}
\usepackage{amsmath,bm}	
\usepackage[ngerman, english]{babel}
\usepackage{braket}

\usepackage{algpseudocode}
\usepackage[colorlinks,bookmarksopen,bookmarksnumbered,allcolors =blue]{hyperref}
\usepackage{tabularx}
\usepackage{booktabs}
\usepackage{xstring}
\usepackage{siunitx}
\usepackage{caption, subcaption}
\usepackage{pgfplots}
\usepackage{pgfplotstable}
\usepackage{cleveref}
\pgfplotsset{compat = newest}
\usepackage{tikz}
\usepackage{tikzscale}
\usepackage{tikz-3dplot}
\usepackage{enumitem}
\usepackage{algorithm}
\usepackage{placeins}

\newcommand{\tensor}[1]{\IfSubStr{ABCDEFGHIJKLMNOPQRSTUVWXYZabcdefghijklmnopqrstuvwxyz}{#1}
        {\mathbf{#1}}
        {\bm{#1}}}

\newcommand{\figref}[1]{Figure~\ref{#1}}

\newcommand{\secref}[1]{Section~\ref{#1}}

\journal{Ultrasonics}

\begin{document}

\newcommand{\hg}[1]{\textcolor{green}{#1}}

\newcommand{\plotting}{1}

\begin{frontmatter}

\title{Reconstructing effective ultrasound transducer models via distributed source inversion}

\author[a]{Tim Bürchner\corref{cor1}}
\ead{tim.buerchner@tum.de}
\cortext[cor1]{Corresponding author}
\author[b]{Simon Schmid}
\author[a,c]{Ernst Rank}
\author[a,d]{Stefan Kollmannsberger}
\author[e]{Andreas Fichtner}

\affiliation[a]{
organization={Chair of Computing in Civil and Building Engineering, Technical University of Munich},
country={Germany}
}

\affiliation[b]{
organization={Chair of Non-destructive Testing, Technical University of Munich},
country={Germany}
}

\affiliation[c]{
organization={Institute of Advanced Study, Technical University of Munich},
country={Germany}
}

\affiliation[d]{
organization={Chair of Data Science in Civil Engineering, Bauhaus University of Weimar},
country={Germany}
}

\affiliation[e]{
organization={Institute of Geophysics, ETH Zurich},
country={Switzerland}
}

\begin{abstract}

Accurate modeling of ultrasound wave propagation is essential for high‑fidelity simulation and imaging in ultrasonic testing.
A primary challenge lies in characterizing the excitation source, particularly for transducers with large apertures relative to the acoustic wavelengths. 
In such cases, non-uniform excitation and spatial interference significantly affect the resulting radiation patterns. 
This paper proposes a distributed source inversion strategy to reconstruct an effective spatio-temporal transducer model that reproduces experimentally measured wavefields. 
The reconstructed source model captures aperture-dependent phase and amplitude variations without the need for detailed knowledge of the transducer structure.
The approach is validated using directivity measurements on an aluminum half-cylinder, where simulations incorporating the reconstructed source model show close agreement with experimental directivity patterns and waveform shapes. 
Finally, synthetic studies on reverse time migration and full-waveform inversion demonstrate that accurate transducer modeling is critical for the success of simulation-based imaging and inversion workflows and significantly improves reconstruction quality.
\end{abstract}


\end{frontmatter}

\section{Introduction}
\label{intro}

Single-element and phased-array ultrasonic testing (UT) are well-established techniques for defect detection and characterization in non-destructive testing (NDT)~\cite{Lempriere2003}. 
A common inspection modality is the B-scan, where a transducer is moved along a line in pulse-echo mode to visualize recorded reflections~\cite{Krautkraemer1980}. 
Although advanced post-processing methods, such as the synthetic aperture focusing technique (SAFT)~\cite{Lorenz1991} and the total focusing method (TFM)~\cite{Holmes2005}, significantly improve image resolution, these approaches assume the validity of the (straight) ray approximation.
Both techniques typically utilize variants of the delay-and-sum algorithm, which primarily account for the kinematic travel times and amplitudes of reflected waves.

Originally developed for seismic exploration, simulation-based imaging frameworks, such as reverse time migration (RTM)~\cite{Baysal1983} and full-waveform inversion (FWI)~\cite{Tarantola1984, Fichtner2011}, aim to exploit the complete information content of recorded wave signals. 
Unlike ray-based methods, these approaches inherently account for complex wave phenomena, including multiple scattering, diffraction, and mode conversion. 
RTM constructs reflectivity images by correlating forward-propagated source wavefields with backward-propagated data wavefields. 
FWI iteratively updates the material model to minimize the misfit between simulated and observed signals. Consequently, the success of these methods depends directly on the accuracy of the underlying wave propagation models.

Over the years, FWI has become a well-established tool in seismic tomography, with applications spanning active-source exploration~\cite{Sheng2006, Sirgue2010}, inversion at regional scales~\cite{Fichtner2009, Fichtner2010}, and at global scales~\cite{Lekic2011, Thrastarson2022}. 
The method has also gained increasing traction in medical ultrasound, particularly in ultrasound computed tomography (USCT) for breast~\cite{Pratt2007, Boehm2022, Ulrich2023} and transcranial brain imaging~\cite{Guasch2020, Marty2022}.
By contrast, the application of FWI in NDT remains comparatively limited. 
This lag is primarily attributed to practical constraints, such as the need for near-real-time results and the frequent inspection of low-cost or high-volume components. 
Nevertheless, certain applications appear well-suited for FWI. 
Recent feasibility studies have explored its use in guided-wave tomography~\cite{Rao2016}, delamination detection in concrete~\cite{Nguyen2017}, material parameter estimation~\cite{Schmid2024}, and phased-array UT~\cite{Buerchner2025}.
Ongoing research tries to further narrow this gap by leveraging GPU hardware in combination with memory-efficient algorithms~\cite{Herrmann2026}. 

Since RTM and FWI directly compare simulated wavefields with measured data, the quality of the reconstruction strongly depends on the accuracy of the numerical wave simulation. Consequently, modeling errors or uncertainties regarding the excitation source can significantly degrade imaging performance. 
While conventional methods like B-scans, SAFT, and TFM are largely agnostic to the source characteristics of the transducer, simulation-based methods require an accurate representation of the emitted ultrasound field. 
In synthetic benchmarks, the source signature is typically prescribed; however, in experimental practice, the transducer response is generally unknown and must be estimated. 
Inaccurate source characterization introduces significant discrepancies between simulation and experiment, which can critically deteriorate the convergence and reliability of simulation-based inversion, as shown in~\cite{Valentine2010} for seismic application.

A straightforward approach for source estimation is the extraction of first-arrival waves from experimental data~\cite{Buerchner2025}. 
More sophisticated strategies, primarily developed for seismic and medical tomography, estimate the source-time function by deconvolving observed waveforms from synthetic data generated using an initial broadband wavelet estimate~\cite{Pratt1999}. 
Groos et al.~\cite{Groos2014} further refined this technique by incorporating frequency-dependent regularization and receiver-specific weighting. 
Since the estimated wavelet is intrinsically coupled to the current material model, it is typically re-computed throughout the FWI workflow. 
This methodology is standard in seismic exploration~\cite{Shin2007, Kalita2017} and has also been applied in ultrasonic NDT~\cite{Nguyen2017, Aktharuzzaman2024}.
In USCT, water-shot calibration measurements are commonly employed to compute the source wavelet~\cite{Ulrich2024}. 
The recovered signal is then associated either with a point source, a distributed source, or a higher-order moment tensor to capture the spatial directivity~\cite{Stump1982}.
Alternative strategies include signature-free or source-independent FWI~\cite{Cheong2006, Choi2011, Tran2012, Tran2013}.
These approaches circumvent the need for source model calibration by eliminating its signature. 
This is achieved by convolving the observed wave data with a reference trace from the modeled wave data, and vice versa.

A particularly notable contribution in the context of USCT is the recent work by Wu et al.~\cite{Wu2023, Wu2025}, who introduced position-dependent weighting for a virtual source array to recover the transducer directivity. 
In their approach, the temporal wavelet is recovered via deconvolution, while the spatial excitation weights are determined through gradient-based optimization. 
Because this calibration is performed using water shots (typically during the initialization phase of the USCT device), analytic solutions of the wave equations can be employed.  
The resulting calibrated virtual source model accurately reproduces the radiation patterns of both synthetic dipole sources and ultrasound transducers in experiments. 
Furthermore, there are approaches tailored to industrial UT where the source wavelets are measured directly.
In~\cite{Schmid2024b} and~\cite{Boxberg2020}, the source time function is recovered by pressing a receiver against the source and measuring the emitted signal. 

Representing the transducer by a single point source is feasible if its spatial extent is considerably smaller than the acoustic wavelengths.
This is the case, for example, for individual elements within a phased array~\cite{Buerchner2025}. 
However, if the dimensions of the transducer are on the order of the minimum wavelength, its spatial characteristics must be taken into account. 
Spatial interference and non-uniform excitation across the aperture (caused by eigenmodes of the piezo crystals~\cite{Eriksson2016, Wiciak2022}) significantly influence the resulting radiation pattern.
Schmid et al.~\cite{Schmid2024} addressed these effects by recovering the characteristics of a disc-shape transducer, specifically its source time function and amplitude distribution using laser Doppler vibrometer (LDV) measurements. 
Overall, the calibrated source model shows good agreement with a directivity measurement on an aluminum half-cylinder.
But, for high aperture angles considerable discrepancies in the wave shapes remain. 

However, not only the amplitude but also the shape of the source time function can fluctuate across the aperture.
Moreover, varying coupling conditions between the transducer and the specimen further influence the emitted field.
These effects are not considered in the approach presented in~\cite{Schmid2024}.
To address these challenges, this paper proposes an adjoint-based distributed source inversion framework that reconstructs an effective spatio-temporal model of an ultrasonic transducer from measurement data.
Similar to finite-source inversions in seismology, our method neither assumes spatial wavelet uniformity nor relies on predefined excitation distributions. 
Instead, it iteratively recovers both the temporal and spatial source characteristics that best reproduce the observed wavefields.
The key contributions of this work are:
\begin{enumerate}[topsep=0pt, noitemsep]
    \item A general and flexible distributed source inversion framework capable of estimating fully spatio-temporal transducer signatures.
    \item Experimental validation using an aluminum half-cylinder dataset, demonstrating that the recovered source model accurately reproduces both waveform shapes and directivity patterns across all aperture angles.
    \item An investigation into the robustness of the proposed approach with respect to source representation and receiver sampling.
    \item A synthetic study on RTM and FWI, illustrating the critical impact of accurate source modeling on the success of simulation-based imaging and inversion.
\end{enumerate}

The remainder of this paper is structured as follows:
\secref{method} introduces the theoretical and methodological background.
\secref{cylinder} applies the distributed source inversion to an aluminum half-cylinder dataset and evaluates the results.
Based on the recovered models, \secref{rtm} investigates the impact of accurate transducer modeling through synthetic RTM and FWI examples.
\secref{discussion} discusses the results and limitations of the proposed approach.
\secref{conclusion} concludes the paper and outlines future research directions.


\section{Methodology}
\label{method}

\label{sec2}
\subsection{Underlying model}

The underlying physics are modeled using the second-order elastic wave equation in a spatial domain $\Omega \subset \mathbb{R}^2$ over a time interval $T = \left[0, t_\mathrm{e}\right]\subset \mathbb{R}$, where $t_\mathrm{e} > 0$ denotes the final time.
Only traction-free boundaries are considered; hence, homogeneous Neumann boundary conditions are imposed on the entire boundary $\partial \Omega$.
Let $\tensor{n}$ denote the outward-pointing normal vector on $\partial \Omega$.
Assuming a linear elastic, isotropic, non-dissipative material, the medium is characterized by the mass density $\rho$ and the fourth-order linear elasticity tensor $\tensor{C}$.
The medium is initially at rest at $t = 0$, and wave propagation is induced by an external force $\tensor{f}$.
The displacement field $\tensor{u}: \Omega \times [0, t_\mathrm{e}] \rightarrow \mathbb{R}^2$ satisfies~\cite{Fichtner2011}
\begin{align}
        &\rho \, \ddot{\tensor{u}} - \nabla \cdot \left( \tensor{C} \colon \nabla \tensor{u} \right) = \tensor{f}
        &&\text{in } \Omega \times (0, t_\mathrm{e}] \label{eq:elastic_we} \\
        &(\tensor{C} \colon \nabla \tensor{u}) \cdot \tensor{n} = \tensor{0}
        &&\text{on } \partial \Omega \times (0, t_\mathrm{e}] \label{eq:material_law}\\
        &\tensor{u} = \tensor{0} 
        &&\text{in } \Omega \text{ at } t = 0 \label{eq:boundary_cond_d}\\ 
        &\dot{\tensor{u}} = \tensor{0} 
        &&\text{in } \Omega \text{ at } t = 0 .
    \label{eq:boundary_cond_n}
\end{align}
These equations constitute the forward model used for simulating wavefields excited by an ultrasonic transducer.
The external force $\tensor{f}$ corresponds to the ultrasonic excitation, which models the piezoelectric transducer as a spatially and temporally distributed source.
Although the presented studies focus on a 2D non-dissipative setting, the proposed framework is general and can be naturally extended to 3D problems, alternative boundary conditions, and dissipative materials.

In the latter examples, measurements are acquired within a single plane.
Accordingly, the domain $\Omega$ is modeled in 2D. 
Aluminum can be considered isotropic, so the stress-strain relation of the linear elastic material is fully determined by two parameters: the Young's modulus $E$ and the Poisson's ratio $\nu$.
The components of the elasticity tensor $\tensor{C}$ are given by
\begin{equation}
    C_{ijkl} = \frac{E \, \nu}{(1 + \nu) \, (1 - 2\nu)} \, \delta_{ij} \, \delta_{kl} + \frac{E}{2 \, (1 + \nu)} \left(\delta_{ik}\, \delta_{jl} + \delta_{il} \, \delta_{jk} \right),
\end{equation}
where $\delta_{ij}$ denotes the Kronecker delta.
The corresponding P- and S-wave speeds of the material are 
\begin{equation}
    v_\mathrm{p} = \sqrt{\frac{E \, (1 - \nu)}{\rho \, (1+\nu) \, (1 - 2\nu)}} \quad \text{and} \quad v_\mathrm{s} = \sqrt{\frac{E}{2\rho \, (1+\nu)}}.
\end{equation}
The formulation supports both compressional and shear wave propagation.

\subsection{Numerical solver}
\label{solver}

All wave simulations in this work were performed using the spectral element method~(SEM)~\cite{Komatitsch1999}, implemented in the parallel solver Salvus~\cite{Afanasiev2019}.
In SEM, Gauss-Lobatto-Legendre (GLL) points serve both as the interpolation nodes of the Lagrange polynomial shape functions and as the quadrature points for numerical integration. 
Following the standard SEM derivation, the spatial discretization of the elastic wave equation leads to a system of second-order ordinary differential equations:
\begin{equation}
    \tensor{M} \ddot{\hat{\tensor{u}}}(t) + \tensor{K} \hat{\tensor{u}}(t) = \tensor{F}(t) 
\end{equation}
where $\tensor{M}$ is the diagonal mass matrix, $\tensor{K}$ the stiffness matrix, $\tensor{F}$ the force vector, and $\hat{\tensor{u}}$ the vector of coefficients corresponding to the SEM basis functions.
Time integration is carried out using the central difference method~(CDM), with a time step size satisfying the Courant–Friedrichs–Lewy~(CFL) condition~\cite{Lewy1928,Ferroni2017}.

\subsection{Optimization problem}

The goal of distributed source inversion~(DSI) is to identify an effective right-hand side $\tensor{f}(\tensor{x}, t)$, such that the modeled wavefield $\tensor{u}$ matches the observed data $\tensor{u}^\mathrm{o}$ within the level of observational noise and measurement uncertainty.
Observations are available only at a finite set of receiver locations $\tensor{x}_r$, $r=1,\dots,N_\mathrm{r}$, where $N_\mathrm{r}$ denotes the number of receivers. 
One seeks to find the optimal source term $\tensor{f}^*$ of the corresponding optimization problem:
\begin{equation}
    \tensor{f}^{*} = \underset{\tensor{f}}{\mathrm{arg\,min}} \, \chi \left(\tensor{f}\right) .
    \label{eq:opt_prob}
\end{equation}
The objective functional $\chi$ quantifies the $L_2$ misfit between simulated and measured data $\tensor{u}^\mathrm{o}$ over all receivers:
\begin{equation}
        \chi \left(\tensor{f}\right) 
        = \frac{1}{2} \sum_{r=1}^{N_\mathrm{r}} \int_T  \left[ w_r(t) \, \left( \tensor{u}(\tensor{f}; \tensor{x}_r, t) - \tensor{u}^\mathrm{o}(\tensor{x}_r, t) \right) \right]^2 \mathrm{d} t
\end{equation}
The receiver-dependent weighting functions $w_r(t)$ allow for emphasizing specific receivers and for applying time-windowing. 
The dependence of the synthetic wavefield on the effective source term $\tensor{f}$ is denoted explicitly.
If the sensor measures the wavefield only in one distinct direction, a directional unit vector~$\tensor{n}_r$ can be introduced that points in the direction of the measurement.

\subsection{Source representation}
\label{source}

To solve the optimization problem the source has to be parameterized.
The effective source is represented by a separable basis of $N_\mathrm{s}$ spatial basis functions $\ell_i$ and $N_\mathrm{t}$ temporal basis functions $k_j$ with coefficients $\hat{\tensor{f}}_{i, j}$:
\begin{equation}
    \tensor{f}(\tensor{x}, t) = \sum_{i=1}^{N_\mathrm{s}} \sum_{j=1}^{N_\mathrm{t}} \, \ell_i(\tensor{x}) \, k_j(t) \, \hat{\tensor{f}}_{i, j} .
\end{equation}
Since many wave solvers, including Salvus, support the definition of point forces, the spatially distributed source is implemented as a collection of point sources at locations $\tensor{x}_i$.
The temporal basis functions are linear interpolants associated with the discrete times $t_j$.
All coefficients $\hat{\tensor{f}}_{i, j}$ are collected in the discrete optimization vector $\hat{\tensor{f}}$.
In principle, the two spatial directions can be excited independently at each source location.
However, if an excitation direction is specified, directional excitation can be modeled by introducing a unit vector $\tensor{n}_{i}$.

\subsection{Optimization algorithm}
\label{sec:algorithm}

The discretized optimization problem is then solved for the calibrated source model $\hat{\tensor{f}}^{*}$.
This work employs the limited-memory Broyden–Fletcher–Goldfarb–Shannon (L-BFGS) algorithm to update the source coefficients.
Algorithm~\ref{alg:lbfgs} summarizes the procedure.
The computation of the gradient and the optimal step length are discussed in \secref{sec:gradient} and \secref{sec:steplength}, respectively.
A detailed description of the L-BFGS approximation and its update rules can be found in standard references such as~\cite{Nocedal2006}.
\begin{algorithm}
\caption{Optimal step length L-BFGS}\label{alg:lbfgs}
\begin{algorithmic}
\Require Initial source model $\hat{\tensor{f}}^{(0)}$, \\
Number of iterations $N_\mathrm{it}$, \\
Number of saved vector pairs $m$
\Ensure Optimized source model $\hat{\tensor{f}}^{*}=\hat{\tensor{f}}^{(N_\mathrm{it})}$
\State Compute initial gradient $\tensor{g}^{(0)} = \nabla \chi\left(\hat{\tensor{f}}^{(0)}\right)$
\For{$k = 0, ..., N_\mathrm{it}-1$}
    \State Compute L-BFGS update direction $\Delta \hat{\tensor{f}}^{(k)}$
    \State Compute optimal step length $\alpha^*$
    \State Update the force vector $\hat{\tensor{f}}^{(k + 1)} = \hat{\tensor{f}}^{(k)} + \alpha^* \Delta \hat{\tensor{f}}^{(k)}$
    \State Compute new gradient $\tensor{g}^{(k+1)} = \nabla \chi\left(\hat{\tensor{f}}^{(k+1)}\right)$
    \State Compute and save $\tensor{s}^{(k)} = \alpha^* \, \Delta \hat{\tensor{f}}^{(k)}$ and $\tensor{y}^{(k)} = \tensor{g}^{(k+1)} - \tensor{g}^{(k)}$
\EndFor
\end{algorithmic}
\end{algorithm}

\subsection{Gradient computation}
\label{sec:gradient}

The following describes the computation of the gradient using the adjoint method.
It starts from the continuous problem.
In the end, the final expressions are discretized. 
This procedure is known as first optimize then discretize~(OTD)~\cite{Fichtner2011}. 
The adjoint method provides an efficient way to compute the gradient of the objective functional with respect to the force $\tensor{f}$, requiring only a single additional adjoint wavefield per forward wavefield. 
The key equations are resumed from~\cite{Fichtner2011}. 
The derivative of the objective functional $\chi$ with respect to the source $\tensor{f}$ due to a perturbation $\delta \tensor{f}$ is given by
\begin{equation}
    \nabla_f \chi \, \delta \tensor{f} = - \int_T \int_\Omega  \tensor{u}^\dagger \cdot \delta \tensor{f} \, \mathrm{d}\Omega \, \mathrm{d}t , \label{eq:continuous_der}
\end{equation}
where adjoint wavefield $\tensor{u}^\dagger: \Omega \times [0, t_\mathrm{e}] \rightarrow \mathbb{R}^2$ satisfies
\begin{align}
    &\rho \, \ddot{\tensor{u}}^\dagger - \nabla \cdot \left( \tensor{C} \colon \nabla \tensor{u}^\dagger \right) = \tensor{f}^\dagger
    &&\text{in } \Omega \times [0, t_\mathrm{e}) \label{eq:elastic_we_adj} \\
    &\tensor{C} \colon \nabla \tensor{u}^\dagger \cdot \tensor{n} = \tensor{0}
    &&\text{on } \partial \Omega \times [0, t_\mathrm{e}) \label{eq:material_law_adk}\\
    &\tensor{u}^\dagger = \tensor{0} 
    &&\text{in } \Omega \text{ at } t = t_\mathrm{e} \label{eq:boundary_cond_d_adj}\\\ 
    &\dot{\tensor{u}}^\dagger = \tensor{0} 
    &&\text{in } \Omega \text{ at } t = t_\mathrm{e} . \label{eq:boundary_cond_n_adj}
\end{align}
Since the dissipation-free elastic wave equation is self-adjoint, the adjoint equation \eqref{eq:elastic_we_adj} has the same form as \eqref{eq:elastic_we} that describes the forward problem, but imposes homogeneous terminal conditions instead of initial conditions.
The adjoint force is given by
\begin{equation}
    \tensor{f}^\dagger = - \sum_{r=1}^{N_\mathrm{r}} w_r^2 \left( \tensor{u}(\tensor{f}; \tensor{x}^r, t) - \tensor{u}^\mathrm{o}(\tensor{x}^r, t) \right) .
\end{equation}
In practice, the adjoint wavefield can be computed with the same numerical methods as the forward wavefield, but is typically integrated backward in time. 
For further details on the adjoint method and its derivation in the context of wave propagation problems, the reader is referred to~\cite{Fichtner2006a, Fichtner2006b, Givoli2021}.

The discrete gradient used in Algorithm~\ref{alg:lbfgs} is equivalent to the derivative of the objective function $\chi$ with respect to the coefficients $\hat{\tensor{f}}_{i, j}$.
According to~\eqref{eq:continuous_der}, the derivative can be computed by evaluating the adjoint solution at position $\tensor{x}_i$ and time $t_j$:
\begin{align}    
\begin{split}
    \frac{\partial \chi}{\partial \hat{\tensor{f}}_{i, j}} &= - \int_T \int_\Omega \tensor{u}^\dagger \, \ell_i (\tensor{x}) \, k_j(t) \, \mathrm{d}\Omega \, \mathrm{d} t \\
    &\approx - \tensor{u}^\dagger(\tensor{x}_i, t_j) \, \Delta t.
\end{split}
\end{align}
where $\Delta t$ is the period between two adjacent points in time.
The gradient vector $\tensor{g}$ comprises all derivatives $\partial \chi / \partial \hat{\tensor{f}}_{i, j}$.

\subsection{Optimal step length} 
\label{sec:steplength}

As stated in \secref{sec:algorithm}, the force vector $\tensor{f}$ is updated iteratively, according to
\begin{equation}
    \tensor{f}^{(k+1)} = \tensor{f}^{(k)} + \alpha \, \Delta \tensor{f}^{(k)}
    \label{eq:update}
\end{equation}
where $k$ denotes the current iteration, $\alpha$ the step length, and $\Delta \tensor{f}$ the update direction.
The optimal step length $\alpha^{*}$ minimizes the objective functional along this direction:
\begin{equation}
    \alpha^{*} = \underset{\alpha}{\mathrm{arg\,min}} \, \chi\left(\tensor{f} + \alpha \, \Delta \tensor{f} \right) .
\end{equation}
Since the wavefield $\tensor{u}$ is linear in $\tensor{f}$, one can write
\begin{equation}
    \chi (\tensor{f} + \alpha \, \Delta \tensor{f} ) 
    = \frac{1}{2} \sum_{r=1}^{N_\mathrm{r}} \int_T  \left[ w_r \, \left( \tensor{u}(\tensor{f}; \tensor{x}^\mathrm{r}) +  \alpha \, \tensor{u}(\Delta\tensor{f}; \tensor{x}^\mathrm{r}) - \tensor{u}^\mathrm{o}(\tensor{x}^\mathrm{r}) \right) \right]^2 \mathrm{d}t .
\end{equation}
Taking the derivative with respect to $\alpha$ gives
\begin{equation}
    \frac{\mathrm{d} \chi}{\mathrm{d} \alpha}
    = \sum_{r=1}^{N_\mathrm{r}} \int_T  w_r^2 \, \left( \tensor{u}(\tensor{f}; \tensor{x}^\mathrm{r}) +  \alpha \, \tensor{u}(\Delta\tensor{f}; \tensor{x}^\mathrm{r}) - \tensor{u}^\mathrm{o}(\tensor{x}^\mathrm{r}) \right)
     \, \tensor{u}(\Delta\tensor{f};\tensor{x}^\mathrm{r}) \, \mathrm{d} t.
\end{equation}
Setting the derivative to zero and solving for $\alpha$ yields the optimal step length
\begin{equation}
    \alpha^* = \frac{ \sum_{r=1}^{N_\mathrm{r}} \int_T w_r^2 \, \left( \tensor{u}^\mathrm{o}(\tensor{x}^\mathrm{r}) - \tensor{u}(\tensor{f};\tensor{x}^\mathrm{r}) \right) \tensor{u}(\Delta\tensor{f}; \tensor{x}^\mathrm{r}) \, \mathrm{d} t }
    {\sum_{r=1}^{N_\mathrm{r}}  \int_T w_r^2 \, (\tensor{u}(\Delta\tensor{f},\tensor{x}^\mathrm{r}))^2 \, \mathrm{d}t }.
    \label{eq:opt_sl}
\end{equation}


\section{Source calibration}
\label{cylinder}

\subsection{Experimental setup and simulation configuration}

\begin{figure}
    \centering
    \begin{tikzpicture}
    \draw[thick, black, fill=white, line width=1.5pt] 
            (-3cm,0) arc (180:360:3cm) -- cycle;
            
    \draw[->, line width=1.5pt] (0,0) -- (3.3cm,0) node[right] {\small$x$};
    \draw[-, line width=.5pt] (0,.6cm) -- (3cm,.6cm) node[midway, above] {\small$300$};
    \draw[-, line width=.5pt] (0,.5cm) -- (0,.7cm);
    \draw[-, line width=.5pt] (3cm,.5cm) -- (3cm,.7cm);
    \draw[->, line width=1.5pt] (0,0) -- (0,-3.5cm) node[right] {\small$y$};

    \node[above] at (-3cm,0) {\small$-90^\circ$}; 
    \node[above] at (3cm,0) {\small$90^\circ$}; 

     \filldraw[black, fill=gray] 
            (-0.1cm,0) rectangle (0.1cm,0.3cm);

    \foreach \angle in {-80, -70, ..., 80}
    {
        \filldraw[black, fill=red, rotate=\angle, shift={(0,-3cm)}]
            (-.05cm,0) rectangle (0.05cm,-0.15cm);
    }
    
    \filldraw[black, fill=gray, shift={(2cm,-3.4cm)}]
            (-0.1cm,0) rectangle (0.1cm,0.3cm);
    \node[right] at (2.05cm,-3.25cm) {\small Source};
    \filldraw[black, fill=red, shift={(2cm,-3.6cm)}]
            (-.05cm,0) rectangle (0.05cm,-0.15cm);
    \node[right] at (2.05cm,-3.675cm) {\small Receiver};

    \draw[-, dashed, line width=1.5pt] (0,0) -- ({sin(30)*3cm},-{cos(30)*3cm});
    \draw[->, black, line width=1.5pt] 
            (0,-1cm) arc (270:300:1cm) node[midway,below] {\small$\theta$};
\end{tikzpicture}    
    \caption{Sketch of half-cylinder experimental setup in $[\si{\milli\meter}]$.}
    \label{fig:half-cyl}
\end{figure}
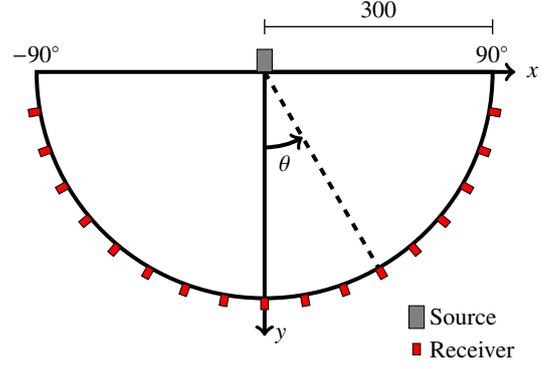

This paper considers ultrasound measurements on an aluminum half-cylinder, a geometry commonly used to characterize the directivity of ultrasonic transducers.
This study utilizes the experimental dataset of~\cite{Schmid2024}, from the setup illustrated in~\figref{fig:half-cyl}.

A single ultrasound transducer was located at the center of the flat surface, while $17$ receivers were attached to the curved surface.
The receiver positions are defined by the angles $\theta_r = - 90^\circ + r \, \Delta \theta$, for $r = 1, ..., 17$ with an angular spacing of $\Delta \theta = 10^\circ$.
Each receiver records the wave component normal to the local surface along measurement directions $\tensor{n}_r = \left[\sin \theta_r, \, \cos \theta_r \right]^\mathrm{T}$.
To improve the signal-to-noise ratio, $100$ measurements were recorded and averaged for each receiver.

The aluminum half-cylinder has a diameter of $\SI{0.3}{\meter}$.
An A V103-RB P-wave transducer with an active diameter of $\SI{12.7}{\milli \meter}$ and a nominal resonant frequency of $\SI{1}{\mega \hertz}$ was employed as the source.
It was excited using a $\SI{150}{\volt}$ square pulse generated by a Trek 2100 HF amplifier together with a TiePie HS5 waveform generator.
Identical transducers were used as receivers.
The material properties of aluminum are: density $\rho = \SI{2707.0}{\kilogram / \meter^3}$, P-wave speed $v_{\mathrm{p}} = \SI{6344.0}{\meter / \second}$ and S-wave speed $v_{\mathrm{s}} = \SI{2887.0}{\meter / \second}$~\cite{Schmid2024}. 

In the numerical model, all wave components up to a maximum frequency of $\SI{2}{\mega \hertz}$ are resolved using approximately $1.5$ quartic elements per minimum wavelength ($\lambda_\mathrm{min} = \SI{1.4435}{\milli \meter}$).
The spatially distributed source is modeled as a collection of point sources acting normal to the flat surface, with source normal vectors $\tensor{n}_i = \left[0, \, 1 \right]^\mathrm{T}$.
The receivers record the wave signal at $\tensor{x}_r$ along their respective surface normals $\tensor{n}_r$. 

\subsection{Distributed source inversion}

The effective distributed source is represented by $20$ point sources uniformly distributed along $x \in [\SI{-0.01}{\meter}, \SI{0.01}{\meter}]$ (including end points), each associated with an independent source wavelet. 
The wavelets are defined within a time window $t \in [0, \SI{0.5e-5}{\second}]$ and tapered using a Tukey window (solid gray line in \figref{fig:windows}). 
Additionally, the wavelets are averaged such that the integral sum of each wavelet is zero.
For the inversion, the P-wave arrivals are isolated from the recorded signals using a second Tukey window applied over $t \in [\SI{2.2e-5}{\second}, \SI{2.9e-5}{\second}]$ (solid blue line in \figref{fig:windows}). 
The experimental data is pre-processed with a Butterworth band-pass filter with cutoff frequencies of $\SI{0.1}{\mega\hertz}$ and $\SI{2.0}{\mega\hertz}$. 
Both simulated and measured waveforms are interpolated onto a reference time line with a sampling rate of $\SI{100}{\mega\hertz}$.
\ifnum\plotting=1\relax
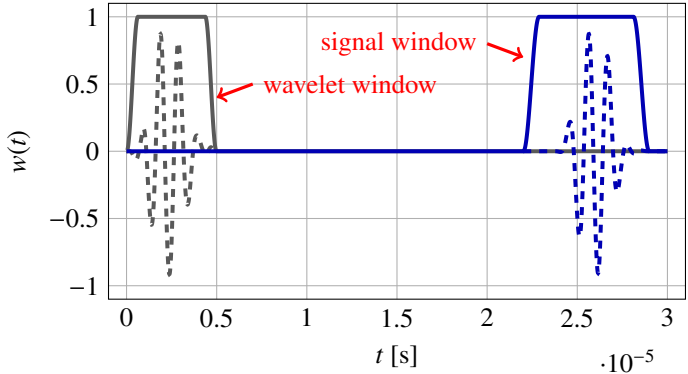
\begin{figure}
    \centering
    \definecolor{TumBlue}{RGB}{0,101,189} 
\definecolor{Orange}{RGB}{227,114,34} 
\definecolor{lightgray204}{RGB}{204,204,204}
\definecolor{darkgray176}{RGB}{176,176,176}
\begin{tikzpicture}
	\begin{axis}[
		xmin = -0.1e-5, xmax = 3.1e-5,
		ymin = -1.1, ymax = 1.1,
        xtick style={color=black},
        x grid style={darkgray176},
        y grid style={darkgray176},
        xmajorgrids,
        ymajorgrids,
        xtick style={draw=none},
        ytick style={draw=none},
		width = .5\textwidth,
		height = .3\textwidth,
		xlabel = {$t$ [\si{\second}]},
		ylabel style={align=center}, 
		ylabel = {$w(t)$},
		]
		
		\addplot[gray!70!black, line width=1.5pt] file[] {tikzpictures/data_cylinder/else/window_wavelet.dat};
        \addplot[
          dashed,
          gray!70!black,
          line width=1.5pt,
        ] table [
            x index = {0}, 
            y expr = \thisrowno{10} * 0.2
        ] {tikzpictures/data_cylinder/no_w/all_wavelets.dat};
        
		\addplot[blue!70!black, line width=1.5pt] file[] {tikzpictures/data_cylinder/else/window_signal.dat};
        \addplot[
          dashed,
          blue!70!black,
          line width=1.5pt,
        ] table [
            x index = {0}, 
            y expr = \thisrowno{1} * 2.5
        ] {tikzpictures/data_cylinder/exp/rec0_exp.dat};
        
        \draw[<-, red, very thick] (axis cs:0.5e-5,0.4) -- (axis cs:0.7e-5,0.5) node[anchor=west] {wavelet window};
        \draw[<-, red, very thick] (axis cs:2.2e-5,0.7) -- (axis cs:2.0e-5, 0.8)  node[anchor=east] {signal window};
	\end{axis}
\end{tikzpicture}%
    \caption{Time windows for wavelets (gray solid) and signals (blue solid) alongside an exemplary source wavelet (gray dashed) and measured signal (blue dashed).}
    \label{fig:windows}
\end{figure}
\fi

Since the transducer predominantly emits P-waves within small aperture angles, the recorded signal amplitudes decrease significantly at larger receiver angles.
To address this imbalance in the data, two DSI weighting schemes are investigated:
\begin{itemize}[label={--}]
    \item DSI 1 (uniform weighting): all receivers contribute equally to the misfit function.
    \item DSI 2 (amplitude-compensating weighting) each receiver is weighted by $1 / \sqrt{A_r}$, where $A_r$ is the maximum amplitude of the recorded signal of receiver $r$.
\end{itemize}
DSI 2 emphasizes the high-aperture regions, encouraging the optimization to better reproduce the waveform characteristics observed at larger angles.
 
The optimization begins from a zero-source model and is carried out in two stages.
First, ten steepest-descent iterations with an optimal step length are performed.
Because the initial updates are large, the resulting inverse Hessian approximation is initially inaccurate.
Subsequently, $40$ iterations of the L-BFGS algorithm are executed following Algorithm~\ref{alg:lbfgs}, with the number of stored vector pairs set to $m = 5$.
\figref{fig:cost} displays the development of the normalized cost for both DSI configurations, the reconstructed source amplitude distributions, and the resulting directivity patterns. 
The kinks in the cost function at iteration ten (annotated with red arrows) mark the transition from steepest descent to L-BFGS.
DSI 1 converges more rapidly to a lower plateau, as the misfit is dominated by high-amplitude, low-aperture receivers. 
Conversely, DSI 2 enforces a balanced fit across all angles, leading to slower but a spatially more uniform convergence.

\ifnum\plotting=1\relax
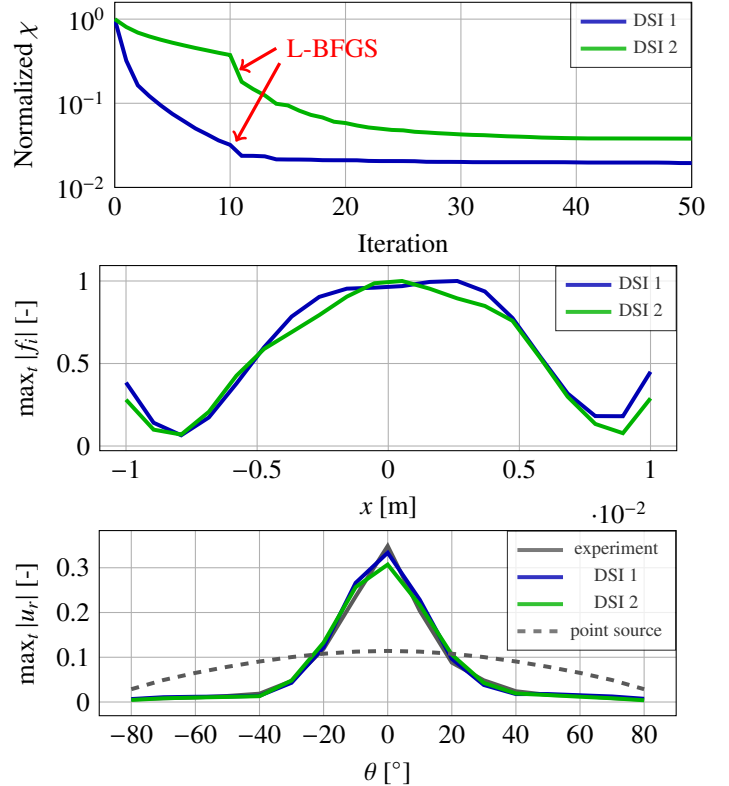
\begin{figure}
    \centering
    \definecolor{TumBlue}{RGB}{0,101,189} 
\definecolor{Orange}{RGB}{227,114,34} 
\definecolor{lightgray204}{RGB}{204,204,204}
\definecolor{darkgray176}{RGB}{176,176,176}
\begin{tikzpicture}
	\begin{semilogyaxis}[
		xmin = 0.0, xmax = 50,
		ymin = 1e-2,
        xtick style={color=black},
        x grid style={darkgray176},
        y grid style={darkgray176},
        xmajorgrids,
        ymajorgrids,
        xtick style={draw=none},
        ytick style={draw=none},
		width = .5\textwidth,
		height = .22\textwidth,
		xlabel = {Iteration},
		ylabel style={align=center}, 
        ylabel = {Normalized $\chi$},
        legend style={font=\scriptsize, opacity = 0.8, at={(1,1)}, anchor=north east},
		]
		
		\addplot[blue!70!black, line width=1.5pt] file[] {tikzpictures/data_cylinder/no_w/obj.dat};
		\addplot[green!70!black, line width=1.5pt] file[] {tikzpictures/data_cylinder/with_w/obj.dat};

        \draw[<-, red, very thick] (axis cs:10.5,3.5e-02) -- (axis cs:14.3,3.0e-01) node[anchor=west] {};
        \draw[<-, red, very thick] (axis cs:10.8,2.5e-01) -- (axis cs:14,4.5e-01) node[anchor=west] {L-BFGS};
        
        \legend{
        DSI 1,
        DSI 2,
        }
	\end{semilogyaxis}
\end{tikzpicture}%
    \definecolor{TumBlue}{RGB}{0,101,189} 
\definecolor{Orange}{RGB}{227,114,34} 
\definecolor{lightgray204}{RGB}{204,204,204}
\definecolor{darkgray176}{RGB}{176,176,176}
\begin{tikzpicture}
	\begin{axis}[
		xmin = -0.011, xmax = 0.011,
        xtick style={color=black},
        x grid style={darkgray176},
        y grid style={darkgray176},
        xmajorgrids,
        ymajorgrids,
        xtick style={draw=none},
        ytick style={draw=none},
		width = .5\textwidth,
		height = .22\textwidth,
		xlabel = {$x$ [\si{\meter}]},
		ylabel style={align=center}, 
        ylabel = {$\max_t \vert f_i \vert$ [-]},
        legend style={font=\scriptsize, opacity = 0.8, at={(1,1)}, anchor=north east},
		]
		
		\addplot[blue!70!black, line width=1.5pt] file[] {tikzpictures/data_cylinder/no_w/source_dist.dat};
		\addplot[green!70!black, line width=1.5pt] file[] {tikzpictures/data_cylinder/with_w/source_dist.dat};
        \legend{
        DSI 1,
        DSI 2,
        }
	\end{axis}
\end{tikzpicture}%
    \definecolor{TumBlue}{RGB}{0,101,189} 
\definecolor{Orange}{RGB}{227,114,34} 
\definecolor{Green}{rgb}{0.01, 0.75, 0.24}
\definecolor{lightgray204}{RGB}{204,204,204}
\definecolor{darkgray176}{RGB}{176,176,176}
\begin{tikzpicture}
	\begin{axis}[
		xmin = -90, xmax = 90,
        xtick style={color=black},
        x grid style={darkgray176},
        y grid style={darkgray176},
        xmajorgrids,
        ymajorgrids,
        xtick style={draw=none},
        ytick style={draw=none},
		width = .5\textwidth,
		height = .22\textwidth,
		xlabel = {$\theta$ [$^\circ$]},
		ylabel style={align=center}, 
        ylabel = {$\max_t \vert u_r \vert$ [-]},
        legend style={font=\scriptsize, opacity = 0.8, at={(1,1)}, anchor=north east},
		]
		
		\addplot[color=gray!70!black, line width=1.5pt] file[] {tikzpictures/data_cylinder/exp/dir_exp.dat};
		\addplot[color=blue!70!black, line width=1.5pt] file[] {tikzpictures/data_cylinder/no_w/dir_sim.dat};
		\addplot[color=green!70!black, line width=1.5pt] file[] {tikzpictures/data_cylinder/with_w/dir_sim.dat};
        
		\addplot[dashed, color=gray!70!black, line width=1.5pt] table[x index = {0}, y expr = \thisrowno{1}] {tikzpictures/data_cylinder/comparison/dir_sim_nS1.dat};
        \legend{
        experiment,
        DSI 1,
        DSI 2,
        point source,
        }
	\end{axis}
\end{tikzpicture}%
    \caption{Development of the cost function (top), normalized source amplitude distributions (middle), and corresponding directivity pattern (bottom).}
    \label{fig:cost}
\end{figure}
\fi

The reconstructed amplitude distribution suggests that the experimental transducer was likely slightly off-centered, as the distribution exhibits a subtle shift to the right. 
Nevertheless, the primary excitation remains within the nominal transducer width of $\SI{12.7}{\milli\meter}$. 
Furthermore, the 2D line-source representation effectively captures the 3D circular aperture effects, i.e., the recovered intensity profile qualitatively matches the expected lateral extent of a circular transducer.
As shown in the bottom panel of \figref{fig:cost}, both configurations successfully recover the experimental directivity pattern in the simulations. 
For reference, the omnidirectional emitting pattern of a point source is visualized.

\figref{fig:wavelets} illustrates the individual wavelet reconstructions for all source points. 
The wavelets vary not only in amplitude but also in shape.
Although the differences of the inverted wavelets for DSI 1 (blue) and DSI 2 (green) are marginal, they significantly impact the waveform shapes at high-aperture angles.
\ifnum\plotting=1\relax
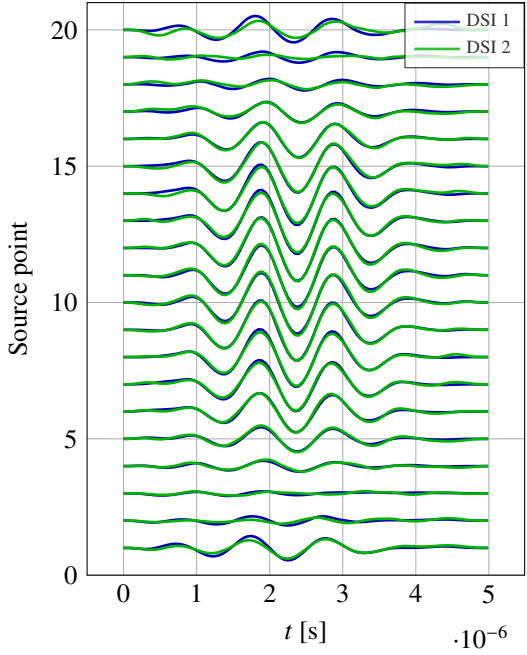
\begin{figure}
    \centering
    \definecolor{TumBlue}{RGB}{0,101,189} 
\definecolor{Orange}{RGB}{227,114,34} 
\definecolor{lightgray204}{RGB}{204,204,204}
\definecolor{darkgray176}{RGB}{176,176,176}
\begin{tikzpicture}
	\begin{axis}[
        xtick style={color=black},
        x grid style={darkgray176},
        y grid style={darkgray176},
        xmajorgrids,
        ymajorgrids,
		ymin = 0, ymax = 21,
        xtick style={draw=none},
        ytick style={draw=none},
		width = .4\textwidth,
		height = 0.5\textwidth,
		xlabel = {$t$ [$\si{\second}$]},
		ylabel style={align=center}, 
        ylabel = {Source point},
        legend style={font=\scriptsize, opacity = 0.8, at={(1,1)}, anchor=north east},
		]

        \foreach \n in {1,...,20}
        {
            \addplot[
              blue!70!black,
              line width=1.0pt,
            ] table [
                x index = {0}, 
                y expr = \n + \thisrowno{\n} * 0.25
            ] {tikzpictures/data_cylinder/no_w/all_wavelets_short.dat};
            
            \addplot[
              green!70!black,
              line width=1.0pt,
            ] table [
                x index = {0}, 
                y expr = \n + \thisrowno{\n} * 0.25
            ] {tikzpictures/data_cylinder/with_w/all_wavelets_short.dat};
        }
        \legend{
            DSI 1,
            DSI 2,
        }
	\end{axis}
\end{tikzpicture}%
    \caption{Reconstructed wavelets for DSI 1 (blue) and DSI 2 (green).}
    \label{fig:wavelets}
\end{figure}
\fi

The resulting P-wave arrivals are shown in \figref{fig:signals} for receiver angles $\theta_r = [-\SI{80}{\degree}, -\SI{60}{\degree}, \ldots, \SI{60}{\degree}, \SI{80}{\degree}]$. 
While DSI~1 accurately matches the amplitudes and waveforms at the central receivers, it struggles to reproduce the signals at larger aperture angles, particularly at $\pm\SI{80}{\degree}$. 
By contrast, DSI~2 achieves good agreement across all angles, accurately reproducing the waveform shapes, albeit with a slight amplitude underestimation at the central receiver.
\ifnum\plotting=1\relax
\begin{figure}
    \centering
    
    \begin{subfigure}{0.23\textwidth}
        \centering
        \definecolor{TumBlue}{RGB}{0,101,189} 
\definecolor{Orange}{RGB}{227,114,34} 
\definecolor{lightgray204}{RGB}{204,204,204}
\definecolor{darkgray176}{RGB}{176,176,176}
\begin{tikzpicture}
	\begin{axis}[
        xtick style={color=black},
        x grid style={darkgray176},
        y grid style={darkgray176},
        xmajorgrids,
        ymajorgrids,
        xtick style={draw=none},
        ytick style={draw=none},
		width = \textwidth,
		height = .8\textwidth,
		xlabel = {$t$ [$\si{\second}$]},
		ylabel style={align=center}, 
		ylabel = {$u_r(t)$ [-]}, 
        legend style={font=\scriptsize, opacity = 0.8, at={(1.1,1)}, anchor=north west},
		]
		
		\addplot[gray!70!black, line width=1.5pt] file[] {tikzpictures/data_cylinder/exp/rec0_exp.dat};
		\addplot[blue!70!black, line width=.75pt] file[] {tikzpictures/data_cylinder/no_w/rec0_sim.dat};
		\addplot[green!70!black, line width=.75pt] file[] {tikzpictures/data_cylinder/with_w/rec0_sim.dat};
        \legend{
        experiment,
        DSI 1,
        DSI 2,
        }
	\end{axis}
    \node[left] at (.65\textwidth, .38\textwidth) {{\color{red}\small$0$}};
\end{tikzpicture}%
    \end{subfigure}
    
    \begin{subfigure}{0.23\textwidth}
        \centering
        \definecolor{TumBlue}{RGB}{0,101,189} 
\definecolor{Orange}{RGB}{227,114,34} 
\definecolor{lightgray204}{RGB}{204,204,204}
\definecolor{darkgray176}{RGB}{176,176,176}
\begin{tikzpicture}
	\begin{axis}[
        xtick style={color=black},
        x grid style={darkgray176},
        y grid style={darkgray176},
        xmajorgrids,
        ymajorgrids,
        xtick style={draw=none},
        ytick style={draw=none},
		width = \textwidth,
		height = .8\textwidth,
		xlabel = {$t$ [$\si{\second}$]},
		ylabel style={align=center}, 
		ylabel = {$u_r(t)$ [-]}, 
		]
		
		\addplot[gray!70!black, line width=1.5pt] file[] {tikzpictures/data_cylinder/exp/rec-20_exp.dat};
		\addplot[blue!70!black, line width=.75pt] file[] {tikzpictures/data_cylinder/no_w/rec-20_sim.dat};
		\addplot[green!70!black, line width=.75pt] file[] {tikzpictures/data_cylinder/with_w/rec-20_sim.dat};
	\end{axis}
    \node[left] at (.65\textwidth, .38\textwidth) {{\color{red}\small$-20^\circ$}};
\end{tikzpicture}%
    \end{subfigure}
    \hfill
    \begin{subfigure}{0.23\textwidth}
        \centering
        \definecolor{TumBlue}{RGB}{0,101,189} 
\definecolor{Orange}{RGB}{227,114,34} 
\definecolor{lightgray204}{RGB}{204,204,204}
\definecolor{darkgray176}{RGB}{176,176,176}
\begin{tikzpicture}
	\begin{axis}[
        xtick style={color=black},
        x grid style={darkgray176},
        y grid style={darkgray176},
        xmajorgrids,
        ymajorgrids,
        xtick style={draw=none},
        ytick style={draw=none},
		width = \textwidth,
		height = .8\textwidth,
		xlabel = {$t$ [$\si{\second}$]},
		ylabel style={align=center}, 
		ylabel = {$u_r(t)$ [-]}, 
		]
		
		\addplot[gray!70!black, line width=1.5pt] file[] {tikzpictures/data_cylinder/exp/rec20_exp.dat};
		\addplot[blue!70!black, line width=.75pt] file[] {tikzpictures/data_cylinder/no_w/rec20_sim.dat};
		\addplot[green!70!black, line width=.75pt] file[] {tikzpictures/data_cylinder/with_w/rec20_sim.dat};
	\end{axis}
    \node[left] at (.65\textwidth, .38\textwidth) {{\color{red}\small$20^\circ$}};
\end{tikzpicture}%
    \end{subfigure}
    
    \begin{subfigure}{0.23\textwidth}
        \centering
        \definecolor{TumBlue}{RGB}{0,101,189} 
\definecolor{Orange}{RGB}{227,114,34} 
\definecolor{lightgray204}{RGB}{204,204,204}
\definecolor{darkgray176}{RGB}{176,176,176}
\begin{tikzpicture}
	\begin{axis}[
        xtick style={color=black},
        x grid style={darkgray176},
        y grid style={darkgray176},
        xmajorgrids,
        ymajorgrids,
        xtick style={draw=none},
        ytick style={draw=none},
		width = \textwidth,
		height = .8\textwidth,
		xlabel = {$t$ [$\si{\second}$]},
		ylabel style={align=center}, 
		ylabel = {$u_r(t)$ [-]}, 
		]
		
		\addplot[gray!70!black, line width=1.5pt] file[] {tikzpictures/data_cylinder/exp/rec-40_exp.dat};
		\addplot[blue!70!black, line width=.75pt] file[] {tikzpictures/data_cylinder/no_w/rec-40_sim.dat};
		\addplot[green!70!black, line width=.75pt] file[] {tikzpictures/data_cylinder/with_w/rec-40_sim.dat};
	\end{axis}
    \node[left] at (.65\textwidth, .38\textwidth) {{\color{red}\small$-40^\circ$}};
\end{tikzpicture}%
    \end{subfigure}
    \hfill
    \begin{subfigure}{0.23\textwidth}
        \centering
        \definecolor{TumBlue}{RGB}{0,101,189} 
\definecolor{Orange}{RGB}{227,114,34} 
\definecolor{lightgray204}{RGB}{204,204,204}
\definecolor{darkgray176}{RGB}{176,176,176}
\begin{tikzpicture}
	\begin{axis}[
        xtick style={color=black},
        x grid style={darkgray176},
        y grid style={darkgray176},
        xmajorgrids,
        ymajorgrids,
        xtick style={draw=none},
        ytick style={draw=none},
		width = \textwidth,
		height = .8\textwidth,
		xlabel = {$t$ [$\si{\second}$]},
		ylabel style={align=center}, 
		ylabel = {$u_r(t)$ [-]}, 
		]
		
		\addplot[gray!70!black, line width=1.5pt] file[] {tikzpictures/data_cylinder/exp/rec40_exp.dat};
		\addplot[blue!70!black, line width=.75pt] file[] {tikzpictures/data_cylinder/no_w/rec40_sim.dat};
		\addplot[green!70!black, line width=.75pt] file[] {tikzpictures/data_cylinder/with_w/rec40_sim.dat};
	\end{axis}
    \node[left] at (.65\textwidth, .38\textwidth) {{\color{red}\small$40^\circ$}};
\end{tikzpicture}%
    \end{subfigure}
    
    \begin{subfigure}{0.23\textwidth}
        \centering
        \definecolor{TumBlue}{RGB}{0,101,189} 
\definecolor{Orange}{RGB}{227,114,34} 
\definecolor{lightgray204}{RGB}{204,204,204}
\definecolor{darkgray176}{RGB}{176,176,176}
\begin{tikzpicture}
	\begin{axis}[
        xtick style={color=black},
        x grid style={darkgray176},
        y grid style={darkgray176},
        xmajorgrids,
        ymajorgrids,
        xtick style={draw=none},
        ytick style={draw=none},
		width = \textwidth,
		height = .8\textwidth,
		xlabel = {$t$ [$\si{\second}$]},
		ylabel style={align=center}, 
		ylabel = {$u_r(t)$ [-]}, 
		]
		
		\addplot[gray!70!black, line width=1.5pt] file[] {tikzpictures/data_cylinder/exp/rec-60_exp.dat};
		\addplot[blue!70!black, line width=.75pt] file[] {tikzpictures/data_cylinder/no_w/rec-60_sim.dat};
		\addplot[green!70!black, line width=.75pt] file[] {tikzpictures/data_cylinder/with_w/rec-60_sim.dat};
	\end{axis}
    \node[left] at (.65\textwidth, .38\textwidth) {{\color{red}\small$-60^\circ$}};
\end{tikzpicture}%
    \end{subfigure}
    \hfill
    \begin{subfigure}{0.23\textwidth}
        \centering
        \definecolor{TumBlue}{RGB}{0,101,189} 
\definecolor{Orange}{RGB}{227,114,34} 
\definecolor{lightgray204}{RGB}{204,204,204}
\definecolor{darkgray176}{RGB}{176,176,176}
\begin{tikzpicture}
	\begin{axis}[
        xtick style={color=black},
        x grid style={darkgray176},
        y grid style={darkgray176},
        xmajorgrids,
        ymajorgrids,
        xtick style={draw=none},
        ytick style={draw=none},
		width = \textwidth,
		height = .8\textwidth,
		xlabel = {$t$ [$\si{\second}$]},
		ylabel style={align=center}, 
		ylabel = {$u_r(t)$ [-]}, 
		]
		
		\addplot[gray!70!black, line width=1.5pt] file[] {tikzpictures/data_cylinder/exp/rec60_exp.dat};
		\addplot[blue!70!black, line width=.75pt] file[] {tikzpictures/data_cylinder/no_w/rec60_sim.dat};
		\addplot[green!70!black, line width=.75pt] file[] {tikzpictures/data_cylinder/with_w/rec60_sim.dat};
	\end{axis}
    \node[left] at (.65\textwidth, .38\textwidth) {{\color{red}\small$60^\circ$}};
\end{tikzpicture}%
    \end{subfigure}
    
    \begin{subfigure}{0.23\textwidth}
        \centering
        \definecolor{TumBlue}{RGB}{0,101,189} 
\definecolor{Orange}{RGB}{227,114,34} 
\definecolor{lightgray204}{RGB}{204,204,204}
\definecolor{darkgray176}{RGB}{176,176,176}
\begin{tikzpicture}
	\begin{axis}[
        xtick style={color=black},
        x grid style={darkgray176},
        y grid style={darkgray176},
        xmajorgrids,
        ymajorgrids,
        xtick style={draw=none},
        ytick style={draw=none},
		width = \textwidth,
		height = .8\textwidth,
		xlabel = {$t$ [$\si{\second}$]},
		ylabel style={align=center}, 
		ylabel = {$u_r(t)$ [-]}, 
		]
		
		\addplot[gray!70!black, line width=1.5pt] file[] {tikzpictures/data_cylinder/exp/rec-80_exp.dat};
		\addplot[blue!70!black, line width=.75pt] file[] {tikzpictures/data_cylinder/no_w/rec-80_sim.dat};
		\addplot[green!70!black, line width=.75pt] file[] {tikzpictures/data_cylinder/with_w/rec-80_sim.dat};
        
	\end{axis}
    \node[left] at (.65\textwidth, .38\textwidth) {{\color{red}\small$-80^\circ$}};
\end{tikzpicture}%
    \end{subfigure}
    \hfill
    \begin{subfigure}{0.23\textwidth}
        \centering
        \definecolor{TumBlue}{RGB}{0,101,189} 
\definecolor{Orange}{RGB}{227,114,34} 
\definecolor{lightgray204}{RGB}{204,204,204}
\definecolor{darkgray176}{RGB}{176,176,176}
\begin{tikzpicture}
	\begin{axis}[
        xtick style={color=black},
        x grid style={darkgray176},
        y grid style={darkgray176},
        xmajorgrids,
        ymajorgrids,
        xtick style={draw=none},
        ytick style={draw=none},
		width = \textwidth,
		height = .8\textwidth,
		xlabel = {$t$ [$\si{\second}$]},
		ylabel style={align=center}, 
		ylabel = {$u_r(t)$ [-]}, 
		]
		
		\addplot[gray!70!black, line width=1.5pt] file[] {tikzpictures/data_cylinder/exp/rec80_exp.dat};
		\addplot[blue!70!black, line width=.75pt] file[] {tikzpictures/data_cylinder/no_w/rec80_sim.dat};
		\addplot[green!70!black, line width=.75pt] file[] {tikzpictures/data_cylinder/with_w/rec80_sim.dat};
	\end{axis}
    \node[left] at (.65\textwidth, .38\textwidth) {{\color{red}\small$80^\circ$}};
\end{tikzpicture}%
    \end{subfigure}
	\caption{Waveforms of the P-wave arrivals using inverted distributed source models. The angle of the corresponding receiver is depicted in red at the top right of each plot.}
    \label{fig:signals}
\end{figure}
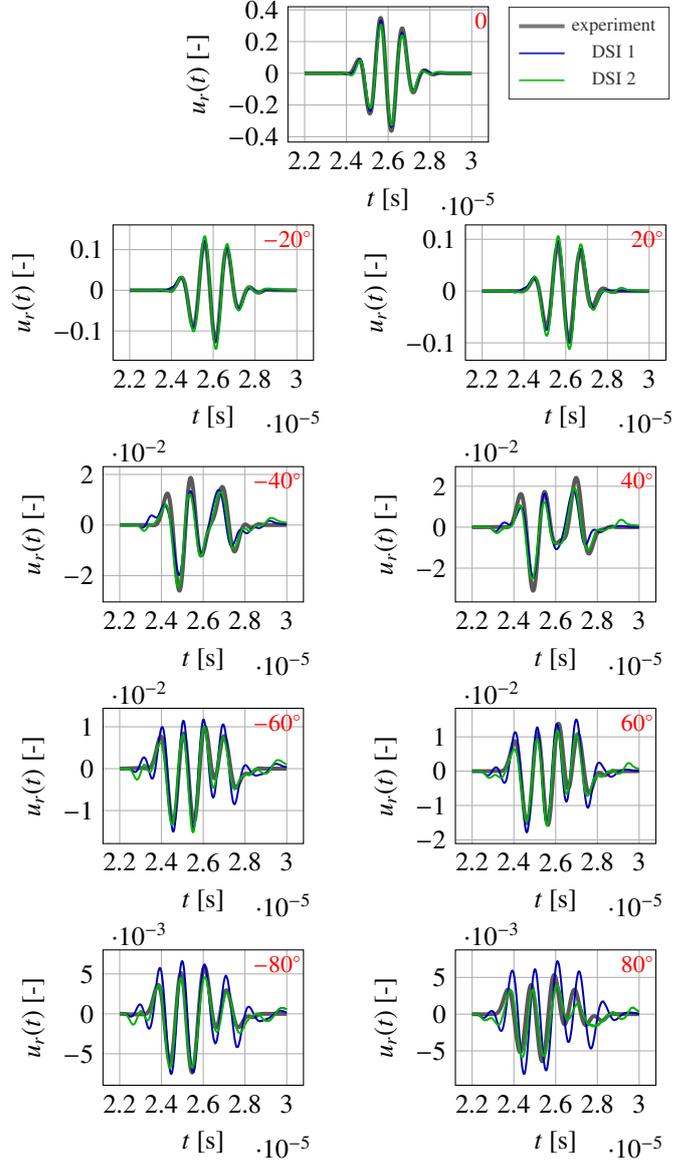
\fi

Finally, \figref{fig:comparison} shows the full wavefields obtained from simulations using a single point source and the DSI 2 model at $t = \SI{2.22e-5}{\second}$. 
While the point source produces a nearly radially uniform P-wavefront, the distributed source generates a pronounced directivity pattern that also influences the S- and Rayleigh-wave components (indicated by red arrows in~\figref{fig:comparison}).
Interference effects across the $\SI{12.7}{\milli\meter}$ aperture (compared to $\lambda_\mathrm{p} = \SI{6.344}{\milli\meter}$ and $\lambda_\mathrm{s} = \SI{2.887}{\milli\meter}$) cause particularly pronounced waveform distortion in tangential directions.
This comparison emphasizes the importance of accurate source modeling for realistic simulations, especially in long-offset configurations, where also the high-aperture content of the wavefield is utilized.

\ifnum\plotting=1\relax
\begin{figure}
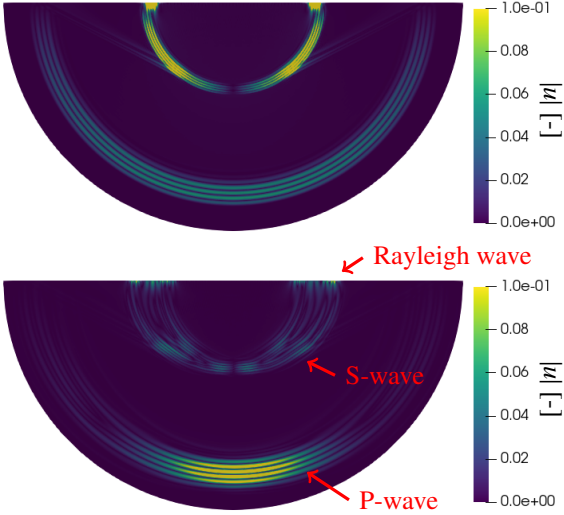

	\centering
	\input{tikzpictures/data_cylinder/wf_1}

    
	\input{tikzpictures/data_cylinder/wf_20}
	\caption{Forward wavefield magnitude from single point source model (top) and reconstructed distributed source model with DSI 2 (bottom) at \mbox{$t = \SI{2.22e-5}{\second}$}.}
	\label{fig:comparison}
\end{figure}
\fi

\subsection{Source sampling}

Additionally, the effect of the spatial source representation on the quality of the reconstructed source model is investigated. 
For the amplitude-compensated weighting (DSI~2), the number of point sources in the distributed source collection is varied as $N_\mathrm{s} = 5,\,10,\,20,\,40,\,80,\,160$. 
An independent inversion is performed for each configuration. 
\figref{fig:study} displays the evolution of the normalized cost functions, the reconstructed source amplitude distributions, and the resulting directivity patterns for all cases.
\ifnum\plotting=1\relax
\begin{figure}
    \centering
    \definecolor{TumBlue}{RGB}{0,101,189} 
\definecolor{Orange}{RGB}{227,114,34} 
\definecolor{lightgray204}{RGB}{204,204,204}
\definecolor{darkgray176}{RGB}{176,176,176}
\begin{tikzpicture}
	\begin{semilogyaxis}[
		xmin = 0.0, xmax = 50,
		ymin = 2e-2,
        xtick style={color=black},
        x grid style={darkgray176},
        y grid style={darkgray176},
        xmajorgrids,
        ymajorgrids,
        xtick style={draw=none},
        xtick style={draw=none},
        ytick style={draw=none},
		width = .5\textwidth,
		height = .22\textwidth,
		xlabel = {Iteration},
		ylabel style={align=center}, 
        ylabel = {Normalized $\chi$},
        legend style={font=\scriptsize, opacity = 0.8, at={(1,1)}, anchor=north east},
		]
		
		\addplot[gray!70!black, line width=1.5pt] file[] {tikzpictures/data_cylinder/study/obj5.dat};
		\addplot[blue!70!black, line width=1.5pt] file[] {tikzpictures/data_cylinder/study/obj10.dat};
		\addplot[green!70!black, line width=1.5pt] file[] {tikzpictures/data_cylinder/study/obj20.dat};
		\addplot[cyan!70!black, line width=1.5pt] file[] {tikzpictures/data_cylinder/study/obj40.dat};
		\addplot[magenta!70!black, line width=1.5pt] file[] {tikzpictures/data_cylinder/study/obj80.dat};
		\addplot[yellow!70!black, line width=1.5pt] file[] {tikzpictures/data_cylinder/study/obj160.dat};
        
        \legend{
        $N_\mathrm{s} = 5$,
        $N_\mathrm{s} = 10$,
        $N_\mathrm{s} = 20$,
        $N_\mathrm{s} = 40$,
        $N_\mathrm{s} = 80$,
        $N_\mathrm{s} = 160$
        }
	\end{semilogyaxis}
\end{tikzpicture}%
    \definecolor{TumBlue}{RGB}{0,101,189} 
\definecolor{Orange}{RGB}{227,114,34} 
\definecolor{lightgray204}{RGB}{204,204,204}
\definecolor{darkgray176}{RGB}{176,176,176}
\begin{tikzpicture}
	\begin{axis}[
		xmin = -0.011, xmax = 0.011,
        xtick style={color=black},
        x grid style={darkgray176},
        y grid style={darkgray176},
        xmajorgrids,
        ymajorgrids,
        xtick style={draw=none},
        ytick style={draw=none},
		width = .5\textwidth,
		height = .22\textwidth,
		xlabel = {$x$ [$\si{\meter}$]},
		ylabel style={align=center}, 
        ylabel = {$\max_t \vert f_i \vert$ [-]},
        legend style={font=\scriptsize, opacity = 0.8, at={(1,1)}, anchor=north east},
		]
		
		\addplot[gray!70!black, line width=1.0pt] file[] {tikzpictures/data_cylinder/study/source_dist5.dat};
		\addplot[blue!70!black, line width=1.0pt] file[] {tikzpictures/data_cylinder/study/source_dist10.dat};
		\addplot[green!70!black, line width=1.0pt] file[] {tikzpictures/data_cylinder/study/source_dist20.dat};
		\addplot[cyan!70!black, line width=1.0pt] file[] {tikzpictures/data_cylinder/study/source_dist40.dat};
		\addplot[magenta!70!black, line width=1.0pt] file[] {tikzpictures/data_cylinder/study/source_dist80.dat};
		\addplot[yellow!70!black, line width=1.0pt] file[] {tikzpictures/data_cylinder/study/source_dist160.dat};
        
	\end{axis}
\end{tikzpicture}%
    \definecolor{TumBlue}{RGB}{0,101,189} 
\definecolor{Orange}{RGB}{227,114,34} 
\definecolor{Green}{rgb}{0.01, 0.75, 0.24}
\definecolor{lightgray204}{RGB}{204,204,204}
\definecolor{darkgray176}{RGB}{176,176,176}
\begin{tikzpicture}
	\begin{axis}[
		xmin = -90, xmax = 90,
        xtick style={color=black},
        x grid style={darkgray176},
        y grid style={darkgray176},
        xmajorgrids,
        ymajorgrids,
        xtick style={draw=none},
        ytick style={draw=none},
		width = .5\textwidth,
		height = .22\textwidth,
		xlabel = {$\theta$ [$^\circ$]},
		ylabel style={align=center}, 
        ylabel = {$\max_t \vert u_r \vert$ [-]},
        legend style={font=\scriptsize, opacity = 0.8, at={(1,1)}, anchor=north east},
		]
		
		\addplot[gray!70!black, line width=1.0pt] file[] {tikzpictures/data_cylinder/study/dir_sim5.dat};
		\addplot[blue!70!black, line width=1.0pt] file[] {tikzpictures/data_cylinder/study/dir_sim10.dat};
		\addplot[green!70!black, line width=1.0pt] file[] {tikzpictures/data_cylinder/study/dir_sim20.dat};
		\addplot[cyan!70!black, line width=1.0pt] file[] {tikzpictures/data_cylinder/study/dir_sim40.dat};
		\addplot[magenta!70!black, line width=1.0pt] file[] {tikzpictures/data_cylinder/study/dir_sim80.dat};
		\addplot[yellow!70!black, line width=1.0pt] file[] {tikzpictures/data_cylinder/study/dir_sim160.dat};
        
	\end{axis}
\end{tikzpicture}%
    \caption{Study with respect to the number of source points: Development of the cost function (top), normalized source amplitude distributions (middle), and corresponding directivity pattern (bottom).}
    \label{fig:study}
\end{figure}
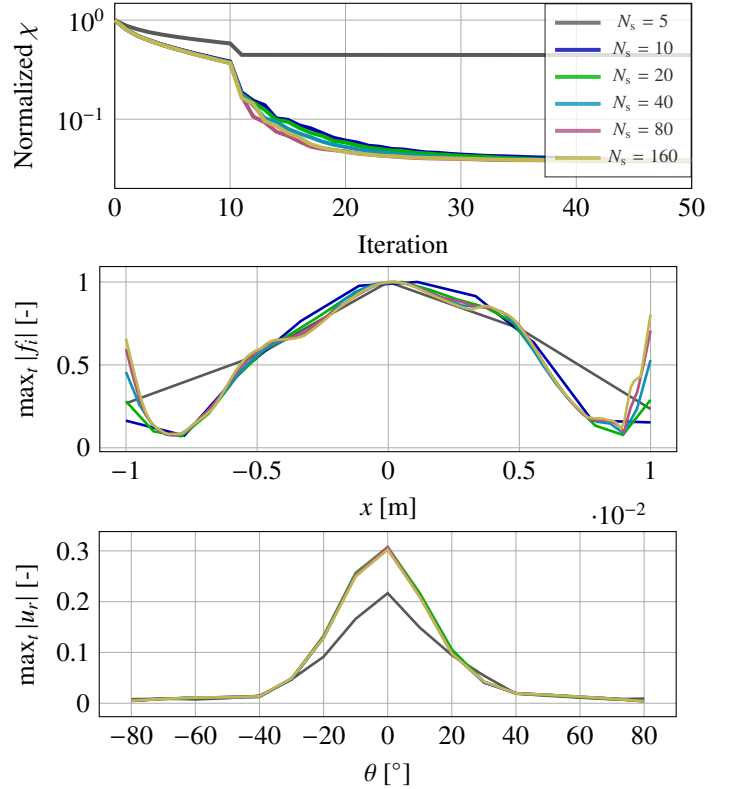
\fi

The results indicate that the number of point sources has only a minor influence on the reconstruction quality, provided a minimum number is met.
For $N_\mathrm{s} = 5$, the inversion fails to recover a suitable source model capable of reproducing the experimental directivity pattern. 
However, once the number of source points is sufficiently dense, only negligible deviations in the reconstructed amplitude distributions and the corresponding directivity patterns are observed.

\subsection{Receiver sampling}

Lastly, the influence of the receiver sampling density is investigated.
DSI2 is performed using four different sets of receivers defined by $\{-80^\circ + \Delta \theta \, r \mid r = 0, 1, ..., N_\mathrm{r}-1\}$,
where the number of receivers used for the inversion is varied as $N_\mathrm{r} = 3, 5, 9, 17$ with corresponding angular spacings $\Delta \theta = \tfrac{160^\circ}{N_\mathrm{r}-1}$.
Independent inversions are carried out for each configuration.
\figref{fig:study2} illustrates the normalized cost functions, the recovered spatial source distributions, and the resulting directivity patterns.

With receiver spacings of $80^\circ$ and $40^\circ$, the reconstructed source model is unable to accurately reproduce the directivity pattern at positions between the receivers used for the inversion.
However, already with a spacing of $20^\circ$ the inverted source model successfully reproduces the observed signals across all receiver positions, including those not included in the optimization.

\ifnum\plotting=1\relax
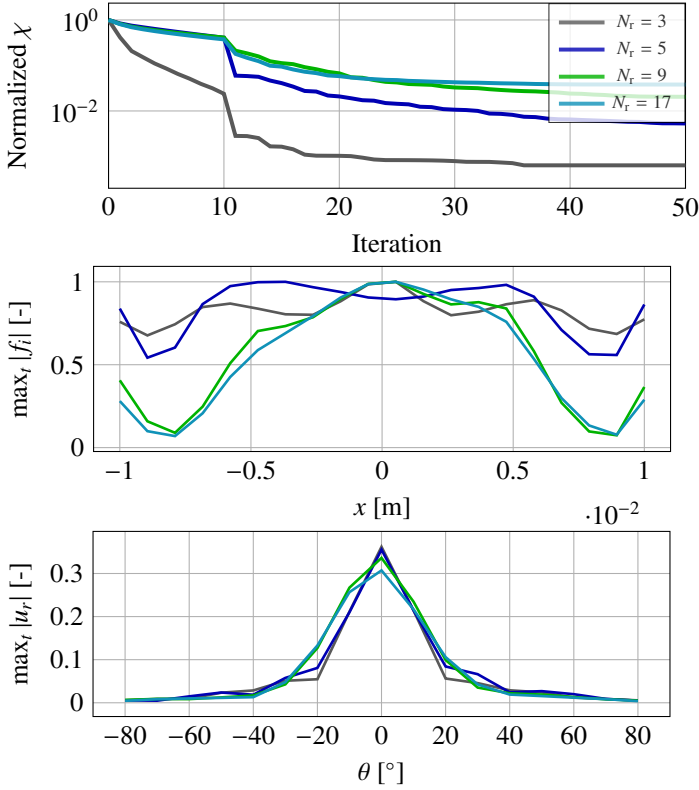
\begin{figure}
    \centering
    \definecolor{TumBlue}{RGB}{0,101,189} 
\definecolor{Orange}{RGB}{227,114,34} 
\definecolor{lightgray204}{RGB}{204,204,204}
\definecolor{darkgray176}{RGB}{176,176,176}
\begin{tikzpicture}
	\begin{semilogyaxis}[
		xmin = 0.0, xmax = 50,
		ymin = 2e-4,
        xtick style={color=black},
        x grid style={darkgray176},
        y grid style={darkgray176},
        xmajorgrids,
        ymajorgrids,
        xtick style={draw=none},
        xtick style={draw=none},
        ytick style={draw=none},
		width = .5\textwidth,
		height = .22\textwidth,
		xlabel = {Iteration},
		ylabel style={align=center}, 
        ylabel = {Normalized $\chi$},
        legend style={font=\scriptsize, opacity = 0.8, at={(1,1)}, anchor=north east},
		]
		\addplot[gray!70!black, line width=1.5pt] file[] {tikzpictures/data_cylinder/study_nr/obj_nr3.dat};
		\addplot[blue!70!black, line width=1.5pt] file[] {tikzpictures/data_cylinder/study_nr/obj_nr5.dat};
		\addplot[green!70!black, line width=1.5pt] file[] {tikzpictures/data_cylinder/study_nr/obj_nr9.dat};
		
		\addplot[cyan!70!black, line width=1.5pt] file[] {tikzpictures/data_cylinder/study/obj20.dat};
        
        \legend{
        $N_\mathrm{r} = 3$,
        $N_\mathrm{r} = 5$,
        $N_\mathrm{r} = 9$,
        $N_\mathrm{r} = 17$,
        }
	\end{semilogyaxis}
\end{tikzpicture}%
    \definecolor{TumBlue}{RGB}{0,101,189} 
\definecolor{Orange}{RGB}{227,114,34} 
\definecolor{lightgray204}{RGB}{204,204,204}
\definecolor{darkgray176}{RGB}{176,176,176}
\begin{tikzpicture}
	\begin{axis}[
		xmin = -0.011, xmax = 0.011,
        xtick style={color=black},
        x grid style={darkgray176},
        y grid style={darkgray176},
        xmajorgrids,
        ymajorgrids,
        xtick style={draw=none},
        ytick style={draw=none},
		width = .5\textwidth,
		height = .22\textwidth,
		xlabel = {$x$ [$\si{\meter}$]},
		ylabel style={align=center}, 
        ylabel = {$\max_t \vert f_i \vert$ [-]},
        legend style={font=\scriptsize, opacity = 0.8, at={(1,1)}, anchor=north east},
		]		
		\addplot[gray!70!black, line width=1.0pt] file[] {tikzpictures/data_cylinder/study_nr/source_dist_nr3.dat};		
		\addplot[blue!70!black, line width=1.0pt] file[] {tikzpictures/data_cylinder/study_nr/source_dist_nr5.dat};
		\addplot[green!70!black, line width=1.0pt] file[] {tikzpictures/data_cylinder/study_nr/source_dist_nr9.dat};
		
		\addplot[cyan!70!black, line width=1.0pt] file[] {tikzpictures/data_cylinder/study/source_dist20.dat};
        
	\end{axis}
\end{tikzpicture}%
    \definecolor{TumBlue}{RGB}{0,101,189} 
\definecolor{Orange}{RGB}{227,114,34} 
\definecolor{Green}{rgb}{0.01, 0.75, 0.24}
\definecolor{lightgray204}{RGB}{204,204,204}
\definecolor{darkgray176}{RGB}{176,176,176}
\begin{tikzpicture}
	\begin{axis}[
		xmin = -90, xmax = 90,
        xtick style={color=black},
        x grid style={darkgray176},
        y grid style={darkgray176},
        xmajorgrids,
        ymajorgrids,
        xtick style={draw=none},
        ytick style={draw=none},
		width = .5\textwidth,
		height = .22\textwidth,
		xlabel = {$\theta$ [$^\circ$]},
		ylabel style={align=center}, 
        ylabel = {$\max_t \vert u_r \vert$ [-]},
        legend style={font=\scriptsize, opacity = 0.8, at={(1,1)}, anchor=north east},
		]		
		\addplot[gray!70!black, line width=1.0pt] file[] {tikzpictures/data_cylinder/study_nr/dir_sim_nr3.dat};
		\addplot[blue!70!black, line width=1.0pt] file[] {tikzpictures/data_cylinder/study_nr/dir_sim_nr5.dat};
		\addplot[green!70!black, line width=1.0pt] file[] {tikzpictures/data_cylinder/study_nr/dir_sim_nr9.dat};
        
		\addplot[cyan!70!black, line width=1.0pt] file[] {tikzpictures/data_cylinder/study/dir_sim20.dat};
        
	\end{axis}
\end{tikzpicture}%
    \caption{Study with respect to the number of receivers: Development of the cost function (top), normalized source amplitude distributions (middle), and corresponding directivity pattern (bottom).}
    \label{fig:study2}
\end{figure}


\fi\section{Impact on RTM and FWI}
\label{rtm}

\subsection{Study setup}

\begin{figure}
    \centering
    \begin{tikzpicture}
    \pgfmathsetmacro{\width}{0.38}
    \pgfmathsetmacro{\height}{0.09}
    \pgfmathsetmacro{\ratio}{\height / \width}
    
    \pgfmathsetmacro{\nS}{9}

    \draw[fill=gray, fill opacity=0.3, draw=white]
    (0.14/\width*.4\textwidth,0) rectangle (0.24/\width*.4\textwidth,0.05/\width*.4\textwidth);

    \draw (0,0) rectangle (.4\textwidth,\ratio*.4\textwidth);

    \draw[fill=white, draw=black] (.2\textwidth, 0.007/\width*0.4\textwidth) circle (0.003/\width*0.4\textwidth);
    \draw[fill=white, draw=black] (.2\textwidth, 0.016/\width*0.4\textwidth) circle (0.002/\width*0.4\textwidth);
    \draw[fill=white, draw=black] (.2\textwidth, 0.0235/\width*0.4\textwidth) circle (0.0015/\width*0.4\textwidth);
    
    \draw[dashed, gray, line width=.5pt] (0.2\textwidth, -0.008\textwidth) -- (0.2\textwidth, 0.11\textwidth);
     
    \draw[<->, black, line width=.2pt] (-0.01\textwidth, 0) -- (-0.01\textwidth, \ratio*0.4\textwidth) node[midway, left] {\scriptsize$90$};
    \draw[-, black, line width=.2pt] (-0.015\textwidth, 0) -- (-0.003\textwidth, 0);
    \draw[-, black, line width=.2pt] (-0.015\textwidth, \ratio*0.4\textwidth) -- (-0.003\textwidth, \ratio*0.4\textwidth);
    
    \draw[<->, black, line width=.2pt] (0,-0.01\textwidth) -- (.4\textwidth, -0.01\textwidth) node[midway, below] {\scriptsize$380$};
    \draw[-, black, line width=.2pt] (0,-0.015\textwidth) -- (0,-0.003\textwidth);
    \draw[-, black, line width=.2pt] (.4\textwidth,-0.015\textwidth) -- (.4\textwidth,-0.003\textwidth);
    
    \draw[<->, black, line width=.2pt] (0.21\textwidth,0) -- (.21\textwidth, 0.007/\width*0.4\textwidth) node[midway, right] {\scriptsize$7$};
    \draw[-, black, line width=.2pt] (0.2\textwidth, 0.007/\width*0.4\textwidth) -- (0.215\textwidth, 0.007/\width*0.4\textwidth);
    
    \draw[<-, black, line width=.2pt] (0.2\textwidth-0.003/\width*.4\textwidth, 0.007/\width*0.4\textwidth) -- (0.135\textwidth, 0.007/\width*0.4\textwidth) node[left] {\scriptsize$\varnothing \, 6$};
    
    \draw[<->, black, line width=.2pt] (0.24\textwidth,0) -- (.24\textwidth, 0.016/\width*0.4\textwidth) node[midway, right] {\scriptsize$16$};
    \draw[-, black, line width=.2pt] (0.2\textwidth, 0.016/\width*0.4\textwidth) -- (0.245\textwidth, 0.016/\width*0.4\textwidth);
    
    \draw[<-, black, line width=.2pt] (0.2\textwidth-0.002/\width*.4\textwidth, 0.016/\width*0.4\textwidth) -- (0.155\textwidth, 0.016/\width*0.4\textwidth) node[left] {\scriptsize$\varnothing \, 4$};
    
    \draw[<->, black, line width=.2pt] (0.27\textwidth,0) -- (.27\textwidth, 0.0235/\width*0.4\textwidth) node[midway, right] {\scriptsize$23.5$};
    \draw[-, black, line width=.2pt] (0.2\textwidth, 0.0235/\width*0.4\textwidth) -- (0.275\textwidth, 0.0235/\width*0.4\textwidth);
    
    \draw[<-, black, line width=.2pt] (0.2\textwidth-0.003/\width*.4\textwidth, 0.0235/\width*0.4\textwidth) -- (0.175\textwidth, 0.0235/\width*0.4\textwidth) node[left] {\scriptsize$\varnothing \, 3$};

    \foreach \i in {0, 1, ..., \nS}
    {
        \filldraw[black, fill=gray] 
        (.04/\width*0.4\textwidth - 0.01/\width*.4\textwidth + \i/\nS*.3/\width*.4\textwidth, \ratio*0.4\textwidth) rectangle 
        (.04/\width*0.4\textwidth + 0.01/\width*.4\textwidth + \i/\nS*.3/\width*.4\textwidth, \ratio*0.4\textwidth + .005 \textwidth);
    }
    \draw[<->, black, line width=.2pt] (.04/\width*0.4\textwidth,\ratio*0.4\textwidth + .015\textwidth) -- (.04/\width*0.4\textwidth+1/\nS*.3/\width*.4\textwidth, \ratio*0.4\textwidth + .015\textwidth) node[midway,above] {\scriptsize$33.3$};    
    \draw[-, black, line width=.2pt] (.04/\width*0.4\textwidth,\ratio*0.4\textwidth + .005\textwidth) -- (.04/\width*0.4\textwidth, \ratio*0.4\textwidth + .02\textwidth);    
    \draw[-, black, line width=.2pt] (.04/\width*0.4\textwidth+1/\nS*.3/\width*.4\textwidth,\ratio*0.4\textwidth + .005\textwidth) -- (.04/\width*0.4\textwidth+1/\nS*.3/\width*.4\textwidth, \ratio*0.4\textwidth + .02\textwidth);

\end{tikzpicture}    
    \caption{Imaging setup adapted from~\cite{Buerchner2025} in $[\si{\milli\meter}]$; the ROI is indicated by the gray shaded area.}
    \label{fig:inversion_example}
\end{figure}
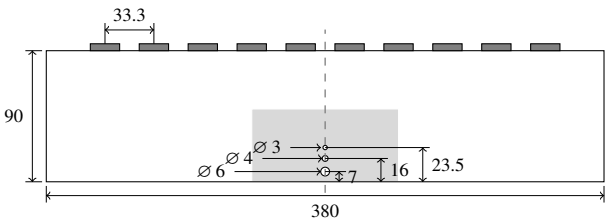

To evaluate the practical significance of accurate source calibration, a synthetic study is conducted. 
The calibrated source model (DSI 2) from the half-cylinder experiment serves as the ground truth to generate synthetic reference data.
The data is then used as input for RTM and FWI.
While RTM produces an image by cross-correlating forward- and backward-propagated wavefields, FWI iteratively updates the material model to minimize the misfit between observed and simulated waveforms.
To exploit the information in the reflection data, the density is inverted in FWI, while the wave speeds are kept constant.
This results in an effective reconstruction of the impedance.
This study adapts the experimental configuration described in \cite{Buerchner2025}. 
Since the original setup utilized a phased array operating at a central frequency of $\SI{2.25}{\mega \hertz}$, the spatial dimensions are doubled to maintain the relative resolution for the $\SI{1}{\mega \hertz}$ transducer. 
\figref{fig:inversion_example} illustrates the resulting setup. 
Three circular holes of increasing size are located close to the back wall. 
Eight distinct transducers are positioned on the top surface with a spacing of $\SI{33.3}{\milli \meter}$, all operating as transmitters and receivers.
The region of interest (ROI), shaded in gray, contains the target features for reconstruction.
For a mathematical derivation of RTM and FWI in the context of ultrasonic NDT, the reader is referred to~\cite{Buerchner2025}.

To investigate the impact of source modeling, the RTM and FWI algorithms are executed using three different source models:
\begin{enumerate}[label=\alph*)]
\item PS (Point Source): A single point source positioned at the center of each transducer location.
\item US (Uniform Source): A distributed source model where a single, identical source wavelet is assigned to all points across the transducer aperture.
\item HS (Heterogeneous Source): The spatially and temporally heterogeneous source model obtained from the DSI 2 calibration.
\end{enumerate}

\subsection{Results}

The imaging and inversion configurations are kept consistent with those used in~\cite{Buerchner2025}. 
In the considered setup, back wall reflections overlap with the reflection and diffraction signals from the defects, leading to significant cross-talk.
As previously demonstrated, excluding signal components associated with the back wall reflections can effectively stabilize the imaging process.

\begin{figure}
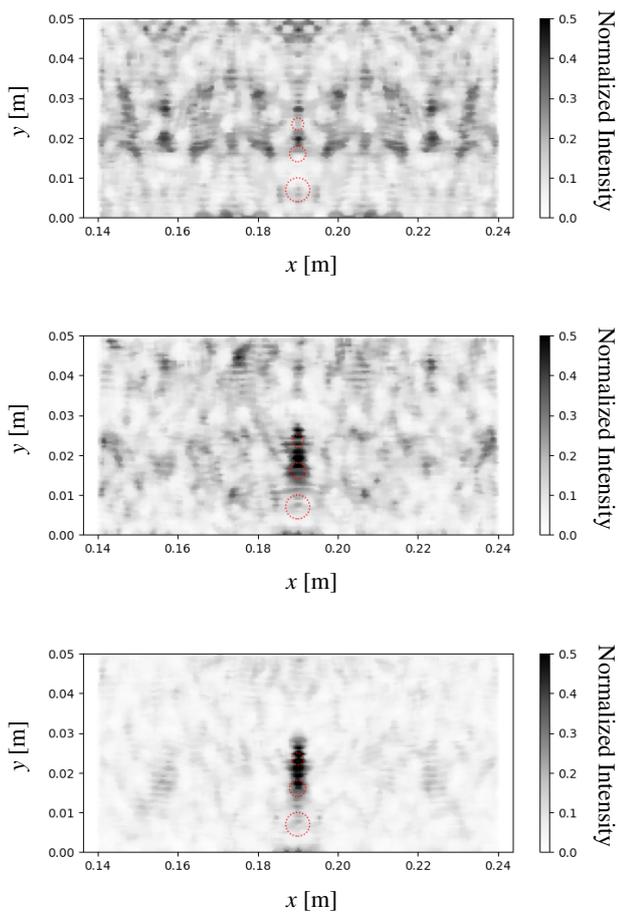

    \centering
    \begin{subfigure}{0.9\linewidth}
        \centering
	    \input{tikzpictures/imaging/rrtm_p}
    \end{subfigure}
    \begin{subfigure}{0.9\linewidth}
        \centering
	    \input{tikzpictures/imaging/rrtm_u}
    \end{subfigure}
    \begin{subfigure}{0.9\linewidth}
        \centering
	    \input{tikzpictures/imaging/rrtm_dsi}
    \end{subfigure}
    \caption{RTM normalized intensity: PS (top); US (middle); HS (bottom).}
    \label{fig:rtm_results}
\end{figure}

For RTM, the direct back wall reflections are excluded from the signals. 
\figref{fig:rtm_results} illustrates the resulting normalized intensity images.
PS fails to identify any of the three holes, as the radiation pattern of the point source differs greatly from the reference simulations. 
By contrast, both US and HS successfully capture the dominant scattering information, namely the direct reflections from the top two holes.
However, RTM remains unable to resolve the individual defects clearly, and the bottom hole remains undetected in all configurations. 
Notably, the HS image exhibits fewer imaging artifacts compared to the US result, indicating that omitting the spatial variations in the source leads to spurious focusing in the cross-correlation of the forward- and backward wavefields.

\begin{figure}
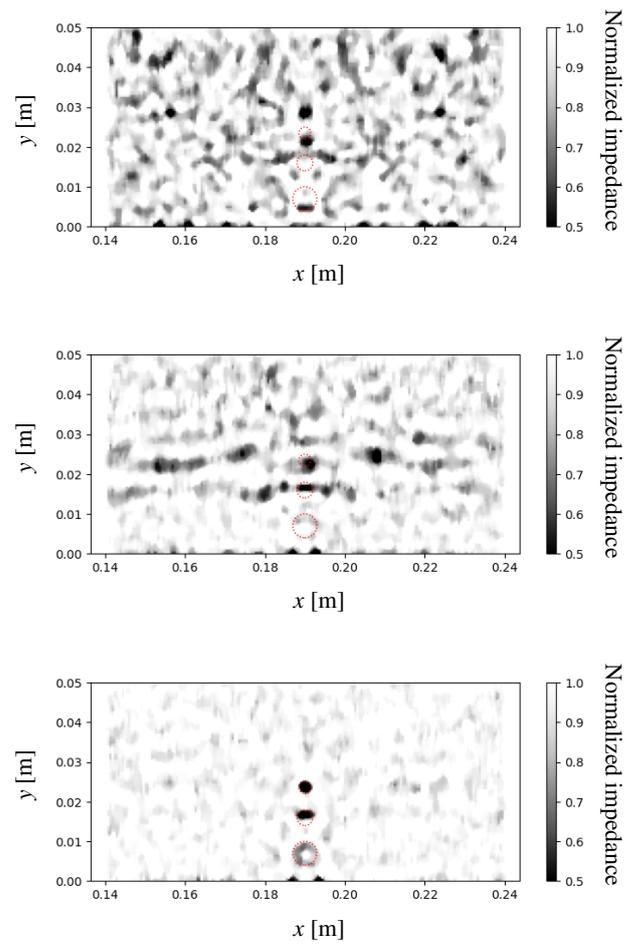

    \centering
    \begin{subfigure}{0.9\linewidth}
        \centering
	    \input{tikzpictures/imaging/rfwi_p}
    \end{subfigure}
    \begin{subfigure}{0.9\linewidth}
        \centering
	    \input{tikzpictures/imaging/rfwi_u}
    \end{subfigure}
    \begin{subfigure}{0.9\linewidth}
        \centering
	    \input{tikzpictures/imaging/rfwi_dsi}
    \end{subfigure}
    \caption{FWI normalized impedance: PS (top); US (middle); HS (bottom).}
    \label{fig:fwi_results}
\end{figure}

To overcome the resolution limits of RTM and recover the third defect, FWI is employed using a two-stage inversion strategy. 
In the first $20$ iterations, the time-domain signals corresponding to the direct P-wave reflections from the back wall are excluded to prioritize the primary scattering from the defects.
Subsequently, the full waveforms are exploited for an additional $20$ iterations to refine the reconstruction and increase axial resolution.
\figref{fig:fwi_results} displays the reconstructed density distributions for the three distinct source models.

Consistent with the RTM results, PS fails to recover any of the holes. 
Furthermore, despite the distributed nature of the US model, it is unable to successfully detect the defects; the mismatch in the radiation pattern results in unreliable gradients, which prevent the optimization from converging toward physically meaningful local minima. 
By contrast, the accurate source modeling in the HS configuration allows the optimization to exploit subtle wavefield components, such as diffracted waves and high-aperture reflections. 
This leads to a significantly improved axial and lateral resolution, enabling the successful reconstruction of all three individual holes.

\section{Discussion}
\label{discussion}

The results presented demonstrate that DSI succeeds in reconstructing effective transducer models that accurately reproduce experimental wavefields in both amplitude and waveform shape. 
The robustness with respect to the spatial discretization of the source representation indicates that the inversion is well-posed and not overly sensitive to the chosen parameterization. 
A sufficient angular density of the receivers is crucial for capturing the spatial complexity of the radiation pattern. 
While a sparse configuration fails to interpolate the directivity correctly, already a $20^\circ$ receiver spacing proves sufficient to resolve the radiation pattern of the transducer in the given setup.
Remaining discrepancies between simulated and measured signals are primarily attributed to the 2D modeling assumptions.

Importantly, the DSI framework reconstructs an effective source rather than explicitly modeling the complex physics of a piezoelectric transducer. 
This provides a pragmatic and flexible approach that implicitly accounts for experimental uncertainties, including transducer placement and coupling conditions. 
Even when detailed transducer specifications are unavailable, the effective source yields realistic excitation characteristics for simulation-based ultrasound approaches. 
Moreover, prior knowledge about the experimental configuration can be easily incorporated through the spatial support of the source and the time-windowing applied to the wavelets and signals.

The subsequent synthetic study underscores the practical necessity of this calibration.
Traditional point source or uniform models lead to artifacts and failed convergence in both RTM and FWI.
When the source models used in the imaging algorithm align with the source characteristics in the reference simulations, even subtle waveform information can be exploited, and closely spaced and near-back-wall defects are accurately identified. 
This confirms that accurate source modeling is a prerequisite for high-resolution, waveform-based imaging.
Finally, the study demonstrates that utilizing single transducers allows the incorporation of high-offset data in contrast to the phased-array configuration in~\cite{Buerchner2025}, thereby improving both the axial and lateral resolution of the reconstruction.

Nevertheless, reconstructing an effective source entails inherent limitations. 
Since the recovered model captures experiment-specific characteristics, its transferability to significantly different configurations is limited; however, it may still provide a valuable initial approximation for similar setups. 
Additionally, the DSI framework 
does not explicitly account for uncertainties in receiver characteristics. 
Further constraints, such as symmetry in the transducer representation or temporal smoothness of the source functions, may be readily incorporated to further stabilize the inversion.


\section{Conclusion}
\label{conclusion}

This work introduced a adjoint-based distributed source inversion (DSI) framework for reconstructing effective transducer models directly from ultrasound measurements.
The approach avoids explicit modeling of piezoelectric elements and instead recovers a spatially distributed, time-dependent source, which implicitly captures experimental uncertainties, such as piezoelectric imperfections, coupling conditions, and slight mispositioning. 
Experimental results demonstrate that the DSI method yields accurate amplitudes and waveforms and is highly robust with respect to the chosen spatial source discretization.

The investigation on receiver sampling highlights that the angular density of measurements is a critical factor.
For the given setup an angular spacing $20^\circ$ was found to be enough to accurately capture and interpolate the directivity of the transducer.
Furthermore, the synthetic imaging study highlights the practical necessity of this calibration framework. 
RTM and FWI using point-source and uniform distributed source approximations fail to recover the investigated defects.
By contrast, accurately representing the source enables to exploit even subtle wavefield components, such as diffraction and high-aperture reflections. 
These components are essential for identifying closely spaced defects near the back wall.

Despite these advantages, the reconstruction of effective source models involves limitations.
The recovered source representation is experiment-dependent and may not transfer directly to different configurations; however, it may serve as a reasonable approximation.
Moreover, the current formulation does not account for uncertainties in receiver positioning or characteristics.

Beyond its use in deterministic simulation-based imaging, DSI has strong potential for supporting machine-learning approaches in ultrasonic testing by providing physically consistent training and validation data. 
By bridging the gap between numerical simulations and experimental reality, DSI provides a practical and flexible tool for realistic wave modeling and advanced waveform-based ultrasonic testing.


\paragraph*{Acknowledgments}
We gratefully acknowledge the funds received by the Deutsche Forschungsgemeinschaft under Grant no. RA 624/29-2 (grant number 438252876), which finance Tim Bürchner.


\begin{thebibliography}{10}

\bibitem{Lempriere2003}
B.~Lempriere, {\em Ultrasound and Elastic Waves: Frequently Asked Questions}.
\newblock Academic Press, 2003.

\bibitem{Krautkraemer1980}
J.~Krautr\"{a}mer and H.~Krautr\"{a}mer, {\em Werkstoffprüfung mit
  Ultraschall}.
\newblock Springer Berlin, Heidelberg, 1980.

\bibitem{Lorenz1991}
M.~Lorenz, L.~van~der Wal, and A.~Berkhout, ``Ultrasonic imaging with
  multi-saft; nondestructive characterization of defects in steel components,''
  {\em Nondestructive Testing and Evaluation}, vol.~6, no.~3, pp.~149--177,
  1991.

\bibitem{Holmes2005}
C.~Holmes, B.~W. Drinkwater, and P.~D. Wilcox, ``Post-processing of the full
  matrix of ultrasonic transmit–receive array data for non-destructive
  evaluation,'' {\em NDT \& E International}, vol.~28, pp.~701--711, 2005.

\bibitem{Baysal1983}
E.~Baysal, D.~D. Kosloff, and J.~W.~C. Sherwood, ``Reverse time migration,''
  {\em GEOPHYSICS}, vol.~48, no.~11, pp.~1514--1524, 1983.

\bibitem{Tarantola1984}
A.~Tarantola, ``Inversion of seismic reflection data in the acoustic
  approximation,'' {\em GEOPHYSICS}, vol.~49, no.~8, pp.~1259--1266, 1984.

\bibitem{Fichtner2011}
A.~Fichtner, {\em Full Seismic Waveform Modelling and Inversion}.
\newblock Advances in Geophysical and Environmental Mechanics and Mathematics,
  Springer Berlin Heidelberg, 2011.

\bibitem{Sheng2006}
J.~Sheng, A.~Leeds, M.~Buddensiek, and G.~T. Schuster, ``Early arrival waveform
  tomography on near-surface refraction data,'' {\em Geophysics}, vol.~71,
  pp.~U47--U57, 07 2006.

\bibitem{Sirgue2010}
L.~Sirgue, O.~Barkved, J.~Dellinger, J.~Etgen, U.~Albertin, and J.~Kommedal,
  ``Thematic set: Full waveform inversion: the next leap forward in imaging at
  valhall,'' {\em First Break}, vol.~28, no.~4, 2010.

\bibitem{Fichtner2009}
A.~Fichtner, B.~L.~N. Kennett, H.~Igel, and H.-P. Bunge, ``{Full seismic
  waveform tomography for upper-mantle structure in the Australasian region
  using adjoint methods},'' {\em Geophysical Journal International}, vol.~179,
  pp.~1703--1725, 12 2009.

\bibitem{Fichtner2010}
A.~Fichtner, B.~L. Kennett, H.~Igel, and H.-P. Bunge, ``Full waveform
  tomography for radially anisotropic structure: New insights into present and
  past states of the australasian upper mantle,'' {\em Earth and Planetary
  Science Letters}, vol.~290, no.~3, pp.~270--280, 2010.

\bibitem{Lekic2011}
V.~Lekić and B.~Romanowicz, ``Inferring upper-mantle structure by full
  waveform tomography with the spectral element method,'' {\em Geophysical
  Journal International}, vol.~185, pp.~799--831, 05 2011.

\bibitem{Thrastarson2022}
S.~Thrastarson, D.-P. van Herwaarden, L.~Krischer, C.~Boehm, M.~van Driel,
  M.~Afanasiev, and A.~Fichtner, ``Data-adaptive global full-waveform
  inversion,'' {\em Geophysical Journal International}, vol.~230,
  pp.~1374--1393, 03 2022.

\bibitem{Pratt2007}
R.~G. Pratt, L.~Huang, N.~Duric, and P.~J. Littrup, ``Sound-speed and
  attenuation imaging of breast tissue using waveform tomography of
  transmission ultrasound data,'' in {\em SPIE Medical Imaging}, 2007.

\bibitem{Boehm2022}
C.~Boehm, L.~Krischer, I.~Ulrich, P.~Marty, M.~Afanasiev, and A.~Fichtner,
  ``Using optimal transport to mitigate cycle-skipping in ultrasound computed
  tomography,'' p.~8, 04 2022.

\bibitem{Ulrich2023}
I.~Ulrich, S.~Noe, C.~Boehm, N.~Korta~Martiartu, B.~Lafci, X.~L. Dean-Ben,
  D.~Razansky, and A.~Fichtner, ``Full-waveform inversion with resolution
  proxies for in-vivo ultrasound computed tomography,'' pp.~1--4, 09 2023.

\bibitem{Guasch2020}
L.~Guasch, O.~Calderon~Agudo, M.-X. Tang, P.~Nachev, and M.~Warner,
  ``Full-waveform inversion imaging of the human brain,'' {\em npj Digital
  Medicine}, vol.~3, 12 2020.

\bibitem{Marty2022}
P.~Marty, C.~Boehm, and A.~Fichtner, ``Elastic full-waveform inversion for
  transcranial ultrasound computed tomography using optimal transport,'' in
  {\em 2022 IEEE International Ultrasonics Symposium (IUS)}, pp.~1--4, 2022.

\bibitem{Rao2016}
J.~Rao, M.~Ratassepp, and Z.~Fan, ``Guided wave tomography based on
  full-waveform inversion,'' {\em IEEE Transactions on Ultrasonics,
  Ferroelectrics, and Frequency Control}, vol.~63, pp.~1--1, 02 2016.

\bibitem{Nguyen2017}
T.~D. Nguyen, K.~T. Tran, and N.~Gucunski, ``Detection of bridge-deck
  delamination using full ultrasonic waveform tomography,'' {\em Journal of
  Infrastructure Systems}, vol.~23, no.~2, p.~04016027, 2017.

\bibitem{Schmid2024}
S.~Schmid, C.~Hachmann, C.~Boehm, L.~Krischer, A.~Mendler, J.~Kollofrath, and
  C.~U. Grosse, ``Estimating young’s moduli based on ultrasound and
  full-waveform inversion,'' {\em Ultrasonics}, vol.~136, p.~107165, 2024.

\bibitem{Buerchner2025}
T.~Bürchner, S.~Schmid, L.~Bergbreiter, E.~Rank, S.~Kollmannsberger, and C.~U.
  Grosse, ``Quantitative comparison of the total focusing method, reverse time
  migration, and full waveform inversion for ultrasonic imaging,'' {\em
  Ultrasonics}, vol.~155, p.~107705, 2025.

\bibitem{Herrmann2026}
L.~Herrmann, T.~Bürchner, L.~Kudela, and S.~Kollmannsberger, ``A
  memory-efficient adjoint method to enable billion parameter optimization on a
  single {{GPU}} in dynamic problems,'' vol.~69, no.~3, p.~52.

\bibitem{Valentine2010}
A.~P. Valentine and J.~H. Woodhouse, ``Reducing errors in seismic tomography:
  combined inversion for sources and structure,'' {\em Geophysical Journal
  International}, vol.~180, pp.~847--857, 02 2010.

\bibitem{Pratt1999}
R.~G. Pratt, ``Seismic waveform inversion in the frequency domain, part 1:
  Theory and verification in a physical scale model,'' {\em GEOPHYSICS},
  vol.~64, no.~3, pp.~888--901, 1999.

\bibitem{Groos2014}
L.~Groos, M.~Schäfer, T.~Forbriger, and T.~Bohlen, ``The role of attenuation
  in 2d full-waveform inversion of shallow-seismic body and rayleigh waves,''
  {\em GEOPHYSICS}, vol.~79, no.~6, pp.~R247--R261, 2014.

\bibitem{Shin2007}
C.~Shin, S.~Pyun, and J.~B. Bednar, ``Comparison of waveform inversion, part 1:
  conventional wavefield vs logarithmic wavefield,'' {\em Geophysical
  Prospecting}, vol.~55, no.~4, pp.~449--464, 2007.

\bibitem{Kalita2017}
M.~Kalita and T.~Alkhalifah, ``Efficient full waveform inversion using the
  excitation representation of the source wavefield,'' {\em Geophysical Journal
  International}, vol.~210, pp.~1581--1594, 05 2017.

\bibitem{Aktharuzzaman2024}
M.~Aktharuzzaman, S.~Anwar, D.~Borisov, and J.~He, ``Experimental full waveform
  inversion for elastic material characterization with accurate transducer
  modeling,'' {\em Mechanical Systems and Signal Processing}, vol.~213,
  p.~111320, 2024.

\bibitem{Ulrich2024}
I.~Ulrich, C.~Boehm, P.~Marty, N.~Korta~Martiartu, B.~Lafci, X.~L. Dean-Ben,
  D.~Razansky, and A.~Fichtner, ``Waveform inversion with calibrated
  source-time functions for improving in-vivo ultrasound computed tomography,''
  p.~15, 04 2024.

\bibitem{Stump1982}
B.~W. Stump and L.~R. Johnson, ``Higher-degree moment tensors — the
  importance of source finiteness and rupture propagation on seismograms,''
  {\em Geophysical Journal International}, vol.~69, pp.~721--743, 06 1982.

\bibitem{Cheong2006}
S.~Cheong, S.~Pyun, and C.~Shin, ``Two efficient steepest-descent algorithms
  for source signature-free waveform inversion,'' {\em Journal of Seismic
  Exploration}, vol.~14, pp.~335--348, 02 2006.

\bibitem{Choi2011}
Y.~Choi and T.~Alkhalifah, ``Source-independent time-domain waveform inversion
  using convolved wavefields: Application to the encoded multisource waveform
  inversion,'' {\em Geophysics}, vol.~76, pp.~125--, 09 2011.

\bibitem{Tran2012}
K.~T. Tran and M.~McVay, ``Site characterization using gauss–newton inversion
  of 2-d full seismic waveform in the time domain,'' {\em Soil Dynamics and
  Earthquake Engineering}, vol.~43, pp.~16--24, 2012.

\bibitem{Tran2013}
K.~T. Tran, M.~McVay, M.~Faraone, and D.~Horhota, ``Sinkhole detection using 2d
  full seismic waveform tomography,'' {\em GEOPHYSICS}, vol.~78, no.~5,
  pp.~R175--R183, 2013.

\bibitem{Wu2023}
X.~Wu, Y.~Li, C.~Su, P.~Li, X.~Wang, and W.~Lin, ``Ultrasound computed
  tomography based on full waveform inversion with source directivity
  calibration,'' {\em Ultrasonics}, vol.~132, p.~107004, 2023.

\bibitem{Wu2025}
X.~Wu, Y.~Li, C.~Su, P.~Li, and W.~Lin, ``Optimal transport assisted full
  waveform inversion for multiparameter imaging of soft tissues in ultrasound
  computed tomography,'' {\em Ultrasonics}, vol.~147, p.~107505, 2025.

\bibitem{Schmid2024b}
S.~Schmid, J.~J. Timothy, E.~Woydich, J.~Kollofrath, and C.~U. Grosse,
  ``Comparison of methods for estimating young's moduli of mortar specimens,''
  {\em Scientific Reports}, vol.~14, p.~14198, Jun 2024.

\bibitem{Boxberg2020}
M.~S. {Boxberg}, M.~{Duda}, K.~{L{\"o}er}, W.~{Friederich}, and J.~{Renner},
  ``{Determining P- and S-wave velocities and Q-values from single ultrasound
  transmission measurements performed on cylindrical rock samples: it's
  possible, when{\textellipsis}},'' in {\em EGU General Assembly Conference
  Abstracts}, EGU General Assembly Conference Abstracts, p.~9178, May 2020.

\bibitem{Eriksson2016}
T.~Eriksson, S.~Ramadas, and S.~Dixon, ``Experimental and simulation
  characterisation of flexural vibration modes in unimorph ultrasound
  transducers,'' {\em Ultrasonics}, vol.~65, pp.~242--248, 2016.

\bibitem{Wiciak2022}
P.~Wiciak, M.~Polak, and G.~Cascante, ``Characterization of ultrasonic
  transducer response using laser doppler interferometer in khz-range for civil
  engineering applications,'' {\em Journal of Nondestructive Evaluation},
  vol.~41, 07 2022.

\bibitem{Komatitsch1999}
D.~Komatitsch and J.~Tromp, ``Introduction to the spectral element method for
  three-dimensional seismic wave propagation,'' {\em Geophysical Journal
  International}, vol.~139, no.~3, pp.~806--822, 1999.

\bibitem{Afanasiev2019}
M.~Afanasiev, C.~Boehm, M.~van Driel, L.~Krischer, M.~Rietmann, D.~A. May,
  M.~G. Knepley, and A.~Fichtner, ``Modular and flexible spectral-element
  waveform modelling in two and three dimensions,'' {\em Geophysical Journal
  International}, vol.~216, no.~3, pp.~1675--1692, 2019.

\bibitem{Lewy1928}
H.~Lewy, K.~Friedrichs, and R.~Courant, ``Über die partiellen
  {D}ifferenzengleichungen der mathematischen {P}hysik,'' {\em Mathematische
  Annalen}, vol.~100, pp.~32--74, 1928.

\bibitem{Ferroni2017}
A.~Ferroni, P.~F. Antonietti, I.~Mazzieri, and A.~Quarteroni,
  ``{Dispersion-dissipation analysis of 3-D continuous and discontinuous
  spectral element methods for the elastodynamics equation},'' {\em Geophysical
  Journal International}, vol.~211, pp.~1554--1574, 09 2017.

\bibitem{Nocedal2006}
J.~Nocedal and S.~Wright, {\em Numerical Optimization}.
\newblock Springer Series in Operations Research and Financial Engineering,
  Springer New York, 2006.

\bibitem{Fichtner2006a}
A.~Fichtner, H.-P. Bunge, and H.~Igel, ``The adjoint method in seismology: I.
  theory,'' {\em Physics of the Earth and Planetary Interiors}, vol.~157,
  no.~1, pp.~86--104, 2006.

\bibitem{Fichtner2006b}
A.~Fichtner, H.-P. Bunge, and H.~Igel, ``The adjoint method in seismology—:
  Ii. applications: traveltimes and sensitivity functionals,'' {\em Physics of
  the Earth and Planetary Interiors}, vol.~157, no.~1, pp.~105--123, 2006.

\bibitem{Givoli2021}
D.~Givoli, ``A tutorial on the adjoint method for inverse problems,'' {\em
  Computer Methods in Applied Mechanics and Engineering}, vol.~380, p.~113810,
  2021.

\end{thebibliography}

\end{document}